\documentclass{aa}
 \usepackage{times}
 \usepackage{graphicx}
 \def\Myr{M$_{\sun}$\,yr$^{-1}$\ }
 \def\Msun{M$_{\sun}$\ }

\begin{document}
 \title{Host Galaxies of Gamma-ray Bursts:\\ Spectral Energy Distributions and
     Internal Extinction.}
 \author{V. V. Sokolov\inst{1,6} \and
      T. A. Fatkhullin\inst{1} \and
      A. J. Castro-Tirado\inst{2,3} \and
      A. S. Fruchter\inst{4} \and
      V. N. Komarova\inst{1,6} \and
      E. R. Kasimova\inst{5} \and \\
      S. N. Dodonov\inst{1} \and
      V. L. Afanasiev\inst{1} \and
      A. V. Moiseev\inst{1}
 }
 \offprints{V. V. Sokolov, \email{sokolov@sao.ru}}

 \institute{Special Astrophysical Observatory of R.A.S.,
         Karachai-Cherkessia, Nizhnij Arkhyz, 369167 Russia
       \and Instituto de Astrof\'{\i}sica de Andaluc\'{\i}a
         (IAA-CSIC), P.O., Box 03004, E-18080 Granada, Spain
       \and Laboratorio de Astrof\'{\i}sica Espacial y F\'{\i}sica Fundamental
         (LAEFF-INTA), P.O., Box 50727, E-28080 Madrid, Spain
       \and Space Telescope Science Institute, 3700 San Martin Drive,
         Baltimore, MD 21218, USA
       \and Institute of Physics, Rostov University,
         Stachki 194, Rostov-on-Don, 344090, Russia
       \and Isaac Newton Institute of Chile, SAO Branch}

 \date{Recieved \today~/~ Submited \today}

 \authorrunning{V. V. Sokolov et al.}

 \titlerunning{GRB host galaxies}

\abstract{
We present $BVR_\mathrm{c}I_\mathrm{c}$ broad-band flux spectra for the
host galaxies of GRB~970508, GRB~980613, GRB~980703, GRB~990123 and
GRB~991208 obtained with the 6-m telescope of SAO RAS.
The comparison of the broad-band flux spectra of these host galaxies with the
template spectral energy distributions (SEDs) of local starburst galaxies of
different morphological types shows that the
$BVR_\mathrm{c}I_\mathrm{c}$ of the hosts
are best fitted by the spectral properties of template SEDs of
starburst galaxies and that there is a significant internal extinction
in these host galaxies.
We derived the absolute magnitudes
of the GRB host galaxies making use of SEDs for the starburst galaxies.
To create theoretical templates we performed the population synthesis modeling
of the continuum spectral
energy distribution of the host galaxies of GRB~970508 and GRB~980703 using
different extinction laws (Cardelli et al. \cite{Card} and
Calzetti et al. \cite{Calz2000}) and
assuming burst and exponential scenarios of star formation.
The comparison of $BVR_\mathrm{c}I_\mathrm{c}$ broad-band flux spectra
with the
local starburst galaxies templates and theoretical templates as well as direct
estimates (using Balmer emission lines) of the internal extinction
shows that it is likely to be of great importance
to take into account effects of the internal extinction in the host galaxies.
From the energy distribution in the spectrum of the host galaxy of
GRB~991208 and from the intensity of their spectral lines
(with allowance for the effects of internal extinction) it
follows that this is a GRB galaxy with the highest
massive star-formation rate of all known GRB
galaxies -- up to hundreds of solar masses per year.
The reduced luminosity of these dusty galaxies (e.g. for the host of GRB~970508
$A_\mathrm{V}\sim 2$ mag, for the host of GRB~980703 $A_\mathrm{V}\sim
0.6$ mag
and for the host of GRB~991208 $A_\mathrm{V}\sim 2$ mag)
could explain the observational fact (it results independently from our
$BVR_\mathrm{c}I_\mathrm{c}$
photometry and from calculated spectral distribution for the subset of galaxies
having been observed with the 6-m telescope):
none of the observed GRB host galaxies with known distances is
brighter than the local galaxies with the luminosity $L_{*}$
(where $L_{*}$ is the ``knee" of the local luminosity function).
\keywords{galaxies: starburst -- galaxies: photometry -- cosmology:
observations -- gamma rays: bursts -- ISM: dust, extintion}
}

\maketitle

\section{Introduction}\label{Intro}
\begin{figure}
\resizebox{\hsize}{!}{\includegraphics[bb=30 30 705 530,clip]{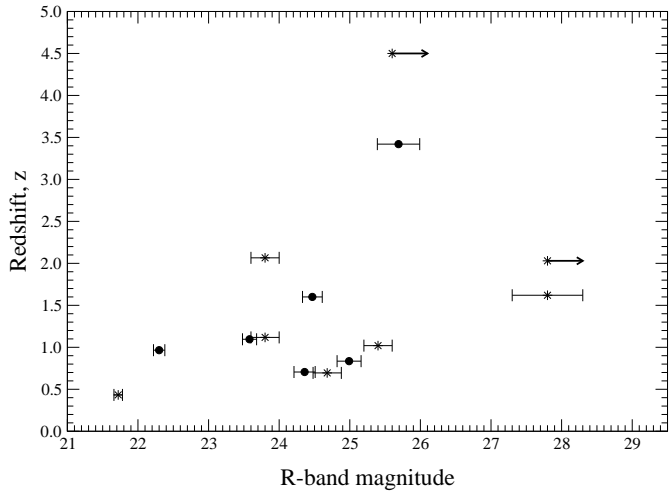}}
\caption{Observed $R$-band magnitude vs. redshift of host
galaxies. The $R$-band magnitudes are corrected for the Galactic extinction.
The $R$-band magnitudes from Table~\ref{phot} are marked with circles
while asterisks correspond to the results of other authors
(Galama et al. \cite{Galama} for the host galaxy of GRB~970228;
Fruchter et al. \cite{Fruchter1999a}; Bloom \cite{Bloom99} for the host
galaxy of GRB~990510;
Hjorth et al. \cite{Hjorth2000} for the host galaxy of GRB990712;
Vreeswijk et al. \cite{Vreeswijk2000a} for the host galaxy of GRB~991216;
Andersen et al. 2000 for the host galaxy of GRB~000131;
Fruchter et al. \cite{Fruchter2000} for the host galaxy of
GRB~000301C;
Metzger et al. \cite{Metzger2000} for the host galaxy of GRB~000418;
Fynbo et al. \cite{Fynbo2000b} for the host galaxy of GRB~000926).
In the case of the host galaxy of GRB~990510, in fact, the $V$-band magnitude
was reported (Bloom \cite{Bloom99}), but using the colour $(V-R)$=0.0 of
the host of
GRB~990123 (see Table~\ref{hosts}, the same redshift), we went from $V$ to $R$
magnitude. The upper limit of brightness of the host galaxy of GRB~000301C is
the magnitude of the optical transient detected by Fruchter et al.
(\cite{Fruchter2000}) on 19 April 2000.}
\label{R-z}
\end{figure}

Currently, more than 20 optical afterglows of long-duration gamma-ray bursts
(GRBs) have been detected (J. Greiner,
http://www.aip.de/$^\sim$jcg/grbgen.html), almost all
within a fraction of an arcsecond of very faint galaxies (Bloom et al.
\cite{Bloom2000a}),
with R-band magnitudes from $\approx$22 to $\approx$28 or more. For example,
in the case of GRB~000301C the host is fainter than $V\approx$ 28.5 mag
(Fruchter et al. \cite{Fruchter2000}) and
we have no certain detection of a host galaxy.
About 14 redshifts have been measured in the range from $z = 0.4$ to
$z = 4.5$ (Castro-Tirado \cite{Alberto2001a}). In Table~\ref{hosts} we present
a summary
of observational data related to the GRB host galaxies. The magnitudes of
the hosts were corrected for the Galactic extinction. Table~\ref{redd} shows
the values of the reddening toward the hosts for the bands
for which we had measurements. In Fig.~\ref{R-z} the $R$-band vs.
redshift diagram for the 14 GRB host galaxies are presented.
\begin{table*}
\caption{Photometry of the 14-th GRB host galaxies with known redshifts}
\label{hosts}
\centering
{\small
\begin{tabular}{lllll}
\hline
\hline
Host       & Date UT         & Band            &  Dereddened      & spectrum \\
	   &                 &                 &  magnitude       &      \\
\hline
GRB~970228 & 01.35 Sep. 1998 & $B$ (VLT-UT1)   &  $>25.11$        & emission host (z=0.6950$\pm$0.0003) \\
	   & 04.71 Sep. 1997 & $V$ (HST/STIS)  &  $25.04\pm 0.20$ & (Bloom et al. 2000) \\
	   & 04.71 Sep.      & $R_\mathrm{c}$ (HST/STIS)&  $24.68\pm 0.20$ &  \\
	   & 04.71 Sep.      & $I_\mathrm{c}$ (HST/STIS)&  $24.38\pm 0.20$ &      \\
	   &                 &                 &                  &      \\
GRB~971214 & 24.85 Jul. 1998 & $V$ (BTA)       &  $25.38\pm 0.3$  & emission host+OT (z=3.418$\pm$0.010)\\
	   & 24.84 Jul.      & $R_\mathrm{c}$ (BTA)     &  $25.65\pm 0.3$  & (Kulkarni~et~al. \cite{Kulkarni})\\
	   &                 &                 &                  & \\
GRB~970508 & 21.74 Aug. 1998 & $B$ (BTA)       &  $25.77\pm 0.19$ & emission host+OT (z=0.8349$pm$0.0003)\\
	   & 23.95 Jul. 1998 & $V$ (BTA)       &  $25.25\pm 0.22$ & emission host \\
	   & 21.74 Aug.      & $R_\mathrm{c}$ (BTA)     &  $24.99\pm 0.17$ & (Metzger~et~al. \cite{Metzger1997},\\
	   & 23.95 Jul.      & $I_\mathrm{c}$ (BTA)     &  $24.07\pm 0.25$ & Bloom~et~al. \cite{Bloom98b})\\
	   &                 &                 &                  &   \\
GRB~980613 & 24.80 Jul. 1998 & $B$ (BTA)       &  $24.77\pm 0.25$ & emission host+OT (z=1.0964$\pm$0.0003)\\
	   & 24.82 Jul.      & $V$ (BTA)       &  $23.94\pm 0.21$ & (Djorgovski~et~al. GCN \#189)\\
	   & 23.00 Jul.      & $R_\mathrm{c}$ (BTA)     &  $23.58\pm 0.1$  & \\
	   &                 &                 &                  & \\
GRB~980703 & 24.05 Jul. 1998 & $B$ (BTA)       &  $23.15\pm 0.12$ & emission host+OT (z=0.9662$\pm$0.0002)\\
	   & 24.06 Jul.      & $V$ (BTA)       &  $22.66\pm 0.10$ & emission  host\\
	   & 24.06 Jul.      & $R_\mathrm{c}$ (BTA)     &  $22.30\pm 0.08$ & (Djorgovski~et~al. \cite{Djorgovski})\\
	   & 24.07 Jul.      & $I_\mathrm{c}$ (BTA)     &  $22.17\pm 0.18$ & \\
	   &                 &                 &                  & \\
GRB~990123 & 8.85 Jul. 1999  & $B$ (BTA)       &  $24.90\pm 0.16$ & absorption host+OT (z=1.6)\\
	   & 8.86 Jul.       & $V$ (BTA)       &  $24.47\pm 0.13$ & (Kelson~et~al. \cite{Kelson}; \\
	   & 8.84 Jul.       & $R_\mathrm{c}$ (BTA)     &  $24.47\pm 0.14$ & Hjorth~et~al. \cite{Hjorth1999})\\
	   & 8.87 Jul.       & $I_\mathrm{c}$ (BTA)     &  $24.06\pm 0.3$  & \\
	   &                 &                 &                  & \\
GRB~990510 & 29   Apr. 2000  & $V$ (HST/STIS)  &  $27.8\pm 0.5$   & absorbtion host+OT (z=1.6187$\pm$0.0015) \\
	   &                 &                 &                  & (Vreeswijk et al. \cite{Vreeswijk2000b})\\
	   &                 &                 &                  &     \\
GRB~990712 & 29 Aug. 1999    & $V$ (HST/STIS)  &  $22.40\pm 0.04$ & emission host+OT (z=0.4337$\pm$0.0004) \\
	   & 29 Aug. 1999    & $R$ (HST/STIS)  &  $21.72\pm 0.06$ & (Hjorth et al. \cite{Hjorth2000}; Vreesvijk et al 2000) \\
	   &                 &                 &                  &   \\
	   &                 &                 &                  & \\
GRB~991208 & 31.90 March 2000& $B$ (BTA)       &  $25.18\pm 0.16$ & emission host+OT \\
	   & 31.84 March     & $V$ (BTA)       &  $24.63\pm 0.16$ & (z=0.7063$\pm$0.0017)\\
	   & 31.96 March     & $R_\mathrm{c}$ (BTA)     &  $24.36\pm 0.15$ & (Dodonov et al. \cite{Dodonov}, GCN \#475)\\
	   & 31.87 March     & $I_\mathrm{c}$ (BTA)     &  $23.70\pm 0.28$ & \\
	   & 04.21 April     & $I_\mathrm{c}$ (NOT)     &  $23.28\pm 0.20$ & \\
	   &                 &                 &                  & \\
GRB~991216 & 17.6 Apr. 2000  & $R$ (HST/STIS)  &  $25.4\pm 0.2$   & absorption host+OT (z=1.02) \\
	   &                 &                 &                  & (Vreeswijk et al. \cite{Vreeswijk1999}, GCN \#496) \\
	   &                 &                 &                  & \\
GRB~000131 & 5.03 March 2000 & $R$ (VLT)       &  $>25.6$         & absorption host+OT (z=4.500$\pm$0.015) \\
	   & 5.02 March      & $I$ (VLT)       &  $>24.7$         & (Andersen et al. \cite{Andersen})\\
	   &                 &                 &                  & \\
GRB~000301C& 19   April 2000 & $V$ (HST/STIS)  &  $>28.3$         & absorpsion host+OT (z=2.0335$\pm$0.0003)\\
	   & 19   April      & $R$ (HST/STIS)  &  $>27.8$         & (Smette et al. 2000, GCN \#603; \\
	   &                 &                 &                  &  S. M. Castro et al. 2000, GCN \#605)\\
	   &                 &                 &                  &  \\
GRB~000418 & 4.17 June 2000  & $R$ (HST/STIS)  &  $23.8\pm 0.2 $  & emission host+OT (z=1.11854$\pm$0.0007)\\
	   &                 &                 &                  & (Bloom et al. \cite{Bloom2000c}, GCN \#661) \\
	   &                 &                 &                  &      \\
GRB~000926 & 27 October 2000 & $R$ (NOT/ALFOSC)&  $23.8\pm 0.2$   & absorption host+OT (z=2.066)\\
	   &                 &                 &                  & (Fynbo et al. \cite{Fynbo2000a}, GCN \#807)\\
\hline
\hline
\end{tabular}
}
\end{table*}

There is a growing suspicion that GRBs are produced during the collapse of
massive stars (see Paczyn\'ski \cite{Pac1999} and  MacFadyen \& Woosley
\cite{MacFadyen} and references therein).
The hosts  of long-duration GRBs are quite blue (Fruchter et al.
\cite{Fruchter2001}),
as one would expect if GRBs were related to star formation,
and the positions of the long-duration GRBs on their hosts are far more
correlated with the light of the galaxy (Bloom et al. \cite{Bloom2000a})
than one would naively expect
if GRBs were formed by the merger of neutron-star binaries, another previously
favored formation scenario (Lattimer \& Schramm \cite{Lattimer};
Paczynski 1986). Additionally, recent spectroscopic
observations of GRB X-ray afterglows have revealed Fe lines in both absorption
and emission (Piro et al. \cite{Piro})
suggesting that the GRBs may even have occured inside a recently formed
supernova remnant.

It is now obvious that the study of host galaxies and their properties
helps in our understanding of the nature of the progenitor systems.
If GRBs are associated with an active star formation, then we might expect the
light of their host galaxies to be affected by internal extinction.  Indeed,
the Balmer line ratios in the spectra of the host of GRB~980703
(Djorgovski et al. \cite{Djorgovski})
and GRB~990712 (Vreeswijk et al. \cite{Vreeswijk2000b}) implied significant
extinction. We pay attention to the fact that in the case of the host
galaxy of GRB~990712 the extinction was
$A_\mathrm{V}=3.4^{+2.4}_{-1.7}$
mag (Vreeswijk et al. \cite{Vreeswijk2000b}). Moreover, the analysis
of the GRB optical afterglow colours shows that the internal extinction
may be sufficient (Castro-Tirado et al. \cite{Alberto1999}; Klose et al.
\cite{Klose} and references therein).
Thus, it is important to take into account the influence of internal
extinction in the interpretation of observational data for estimates of
star formation rate (SFR).

Our aim in this paper is to estimate the effects of internal
extinction on the host galaxy parameters SFR and luminosity.
We use our photometric observational data on GRB host galaxies,
obtained with the 6-m telescope during the period 1998--2000 
and other, already available data (Bloom et al. \cite{Bloom98b};
Djorgovski et al. \cite{Djorgovski}).
We performed a population synthesis modeling for two host
galaxies (GRB~970508 and GRB~980703), using our
$BVR_\mathrm{c}I_\mathrm{c}$ and published near-infrared
data from Pian et al. (\cite{Pian}) and Bloom et al. (\cite{Bloom98a}),
to create a set of theoretical template SEDs for
more exact estimation of reddening in the host galaxies.

Sect.~\ref{Obs-reduc} describes our
observational data. In Sect.~\ref{SAO-Keck} we compare our
$BVR_\mathrm{c}I_\mathrm{c}$ broad-band
flux spectra to the spectra of the GRB~970508  (Bloom et al. \cite{Bloom98b})
and GRB~980703  (Djorgovski et al. \cite{Djorgovski}) host galaxies.
Using the comparison of our $BVR_\mathrm{c}I_\mathrm{c}$ data with
template SEDs of local star-forming
galaxies with different internal reddening, we roughly estimate the extinction
in host galaxies ({\it a first approximation}). The results are given in
Sect.~\ref{BVRI-SB}.
This comparison allows us to estimate the $K$-correction for the
hosts. We used the template SEDs from Sect.~\ref{BVRI-SB} to calculate
the $K$-correction as well as other methods for the estimates of absolute
magnitudes of the host galaxies in Sect.~\ref{Est-Mabs}.
We describe and perform a theoretical modeling in Sect.~\ref{Modelling}.
This modeling allows us to make more exact estimates of the internal extinction
(in {\it a second approximation})
using rough estimates from Sect.~\ref{BVRI-SB}.
The results of the comparison of theoretical templates with our
$BVR_\mathrm{c}I_\mathrm{c}$
and infrared data of the host galaxies of GRB~970508 and GRB~980703 with
essentialy different luminosities
are used to estimate the effects of the internal
extinction in the host galaxies.
In Sect.~\ref{Est_SFR} we make estimates of SFR
for the host galaxy of GRB~991208 using different methods.
The final discussion and conlusions are given in Sect.~\ref{Disc} and
Sect.~\ref{Concl}.

\section{Observations and data reduction}\label{Obs-reduc}
Observations of the GRB host galaxies were performed during July 1998-March 2000
(see the log of observations in Table~\ref{phot}). The primary focus CCD
photometer of the 6-m telescope of SAO RAS
with the standard (Johnson-Kron-Cousins) photometric
$BVR_\mathrm{c}I_\mathrm{c}$ system was used.
The photometric calibrations were performed
using the Landolt standard stars (Landolt \cite{Landolt}).
All data were processed in the standard manner: bias and dark corrected,
flat-fielded and inspected for cosmic-rays. The photometry of host
galaxies was performed
with a circular aperture fixed in all bands for each host galaxy.
The ESO-MIDAS\footnote{MIDAS (Munich Image Data Analysis System) is
distributed and supported by European South Observatory.
http://www.eso.org/progects/esomidas/}
package was used for the data processing and photometry.
\begin{table}
\centering
\caption{Galactic reddening toward host galaxies}
\label{redd}
\begin{tabular}{lllll}
\hline
\hline
Host       & $A_\mathrm{B}$  &  $A_\mathrm{V}$ & $A_\mathrm{R}$  & $A_\mathrm{I}$ \\
\hline
GRB~970228 & 0.97   & 0.73   & 0.54   & 0.35  \\
GRB~970508 & 0.12   & 0.09   & 0.07   & 0.04  \\
GRB~971214 & 0.07   & 0.05   & 0.04   & 0.02  \\
GRB~980613 & 0.36   & 0.28   & 0.22   & 0.13  \\
GRB~980703 & 0.25   & 0.19   & 0.14   & 0.09  \\
GRB~990123 & 0.07   & 0.05   & 0.04   & 0.02  \\
GRB~990510 &        & 0.7    &        &       \\
GRB~990712 &        & 0.11   & 0.08   &       \\
GRB~991208 & 0.07   & 0.05   & 0.04   & 0.02  \\
GRB~991216 &        &        & 1.48   &       \\
GRB~000131 &        & 0.13   & 0.08   &       \\
GRB~000301C&        & 0.16   & 0.12   &       \\
GRB~000418 &        &        & 0.09   &       \\
GRB~000926 &        &        & 0.06   &       \\
\hline
\hline
\end{tabular}
\end{table}
Table \ref{phot} shows the results of our photometry for the galaxies
with more than two filters.
To tranform the magnitudes (see Table~\ref{hosts}) into the flux densities
we have used zero-points from Fukugita et al. (\cite{Fukugita}).
All these flux densities are corrected for the Galactic extinction.

In all cases our $BVR_\mathrm{c}I_\mathrm{c}$ observations were performed
more than
20 days after the GRB occurrence (see Table~\ref{phot}) and the contribution of
the OT to the host galaxy continuum flux is expected to be minimal.

\begin{table*}
\caption{The $BVR_\mathrm{c}I_\mathrm{c}$ photometrical observations of
the 6-th GRB host galaxies with the 6-m telescope}
\label{phot}
\centering
{\small
\begin{tabular}{lllcll}
\hline
\hline
Host       & Date UT         & Band  & Exp. & Seeing     & flux density, $\mu$Jy \\
	   &                 &       & (s)  &            &\\
\hline
GRB~971214 & 24.85 Jul. 1998 & $V$   & 600  & $1\farcs2$ & $0.25^{+0.08}_{-0.06}$\\
	   & 24.84 Jul.      & $R_\mathrm{c}$ & 600  & $1\farcs2$ & $0.17^{+0.05}_{-0.04}$\\
	   &                 &       &      &            &\\
GRB~970508 & 21.74 Aug. 1998 & $B$   & 4200 & $1\farcs3$ & $0.20^{+0.04}_{-0.03}$\\
	   & 23.95 Jul. 1998 & $V$   & 2000 & $1\farcs3$ & $0.28^{+0.07}_{-0.05}$\\
	   & 21.74 Aug.      & $R_\mathrm{c}$ & 3000 & $1\farcs3$ & $0.30^{+0.05}_{-0.04}$\\
	   & 23.95 Jul.      & $I_\mathrm{c}$ & 2000 & $1\farcs3$ & $0.56^{+0.15}_{-0.11}$\\
	   &                 &       &      &            &\\
GRB~980613 & 24.80 Jul. 1998 & $B$   & 700  & $1\farcs3$ & $0.50^{+0.13}_{-0.10}$\\
	   & 24.82 Jul.      & $V$   & 600  & $1\farcs3$ & $0.95^{+0.20}_{-0.17}$\\
	   & 23.00 Jul.      & $R_\mathrm{c}$ & 1800 & $1\farcs5$ & $1.12^{+0.11}_{-0.10}$\\
	   &                 &       &      &            &\\
GRB~980703 & 24.05 Jul. 1998 & $B$   & 480  & $1\farcs3$ & $2.19^{+0.25}_{-0.23}$\\
	   & 24.06 Jul.      & $V$   & 320  & $1\farcs2$ & $3.09^{+0.30}_{-0.27}$\\
	   & 24.06 Jul.      & $R_\mathrm{c}$ & 300  & $1\farcs2$ & $3.63^{+0.26}_{-0.24}$\\
	   & 24.07 Jul.      & $I_\mathrm{c}$ & 360  & $1\farcs2$ & $3.31^{+0.58}_{-0.49}$\\
	   &                 &       &      &            &\\
GRB~990123 & 8.85 Jul. 1999  & $B$   & 600  & $1\farcs5$ & $0.44^{+0.06}_{-0.06}$\\
	   & 8.86 Jul.       & $V$   & 600  & $1\farcs3$ & $0.59^{+0.07}_{-0.07}$\\
	   & 8.84 Jul.       & $R_\mathrm{c}$ & 600  & $1\farcs1$ & $0.48^{+0.07}_{-0.06}$\\
	   & 8.87 Jul.       & $I_\mathrm{c}$ & 600  & $1\farcs3$ & $0.56^{+0.18}_{-0.13}$\\
	   &                 &       &      &            &\\
GRB~991208 & 31.90 March 2000& $B$   & 1795 & $3\farcs0$ & $0.32^{+0.05}_{-0.04}$\\
	   & 31.84 March     & $V$   & 1490 & $2\farcs1$ & $0.48^{+0.08}_{-0.07}$\\
	   & 31.96 March     & $R_\mathrm{c}$ & 1260 & $2\farcs1$ & $0.52^{+0.08}_{-0.07}$\\
	   & 31.87 March     & $I_\mathrm{c}$ & 360  & $2\farcs6$ & $0.77^{+0.23}_{-0.18}$\\
	   & 04.21 April$^*$ & $I_\mathrm{c}$ & 3800 & $1\farcs2$ & $1.17^{+0.24}_{-0.20}$\\
\hline
\hline
\end{tabular}
}
\parbox[t]{0.6\textwidth}{
$^*$ The observations were obtained with the 2.5-m NOT (Castro-Tirado
et al. \cite{Alberto2001b})
}
\end{table*}

\section{Comparison of $BVR_\mathrm{c}I_\mathrm{c}$ with spectra of host
galaxies}\label{SAO-Keck}
Usually the spectroscopy of host galaxies is carried out
when there is a significant contamination of the OT flux
(see the column ``spectrum" in Table~\ref{hosts}).
Here we present photometry of pure host galaxies or with a negligible
OT contribution.
In some cases (e.g. as for GRB~990123),
absorption spectra were obtained. There is a range of
redshifts for which  no strong emission lines ([\ion{O}{ii}]\,3727\AA,
[{\ion{O}{iii}]\,4959,5007\AA\ and Balmer lines) are observed in the
optical domain.
In that case we can infer a lower limit to the redshift and it adds
some uncertainty to the determination of the parameters of host galaxies.
In the case of emission spectra, the luminosity (i.e. the equivalent width of
the lines) is important. However,
in the case of a sufficiently bright OT it is possible that the OT continuum
would affect these parameters.

Nevertheless, Table~\ref{hosts} shows
two cases of pure host galaxy spectra: GRB~970508 (Bloom et al. \cite{Bloom98b})
and GRB~980703 (Djorgovski et al. \cite{Djorgovski}) which allow us
to perform further theoretical modeling of the continuum spectral energy
distribution of these hosts, making use of these spectra and our
$BVR_\mathrm{c}I_\mathrm{c}$
photometry (see Sect.~\ref{Modelling}).
Here we show these two examples to demonstrate that
$BVR_\mathrm{c}I_\mathrm{c}$ photometrical
spectral distribution describes well the spectral continua of host galaxies
in the absence of any broad spectral features. Furthermore, in the case of
absence of the spectra of a host galaxy itself (without OT), the
$BVR_\mathrm{c}I_\mathrm{c}$
photometry is the only tool to study spectral energy distribution for
these faint objects.

As can be seen from
the comparison of our $BVR_\mathrm{c}I_\mathrm{c}$ broad-band photometry
with the
spectra of these host galaxies (see Fig.~\ref{grb970508Keck_SAO} and
\ref{grb980703Keck_SAO}), our photometric points are in perfect agreement
with the continua of the host galaxies. Moreover, the errors in the
$BVR_\mathrm{c}I_\mathrm{c}$
photometry are significantly smaller than the noise present in the spectra
(at least
for the host of GRB~970508, which has typical magnitudes of GRB host
galaxies).
Thus, we can conclude that our $BVR_\mathrm{c}I_\mathrm{c}$ broadband
flux spectra
approximate the spectral energy distributions (SEDs) properly
at least in the cases where there are no broad spectral features in the
spectra of these galaxies. This is the case of the pure spectra
of the GRB~970508 and GRB~980703 hosts.

Additionaly, Fig.~\ref{grb980703Keck_SAO} shows
the time evolution of the OT contribution
to the spectrum of the host galaxy of GRB~980703.
As we can see from the spectra obtained by Djorgovski et al. (\cite{Djorgovski})
in {\it two epochs: 07 July 1998 UT and 19 July 1998 UT,} the central intensities
of Balmer lines H$\gamma$ and H$\delta$ remain close to constant during
the observations. To explain this, it can be considered that
the OT shone through an \ion{H}{ii} region displaying Balmer lines in absorption,
which may be evidence for a direct link between GRBs and
star-forming regions, where massive O and B stars ionize the interstellar
medium. The assosiation with \ion{H}{ii} regions implies
that long-duration GRBs are closely connected with the massive star formation
and, consequently, with some explosions of Type Ib/c core collapse Supernovae
(Bartunov et al. \cite{Bartunov}).

In fact, we consider that broad-band photometry
remains a useful tool for
studying host galaxies, mainly due to the fact that it is impossible
to obtain a spectrum for each GRB galaxy.
Indeed, spectroscopy of extended objects of the integral 25-26th magnitude
and fainter is a nontrivial task, even for 8-10\,m
class ground-based telescopes, because there is a background night-sky light
which begins to dominate the target spectrum for
the long time exposures needed to achieve a good signal-to-noise ratio.

\begin{figure*}[th]
\centering
\includegraphics[width=17cm,bb=55 175 540 475,clip]{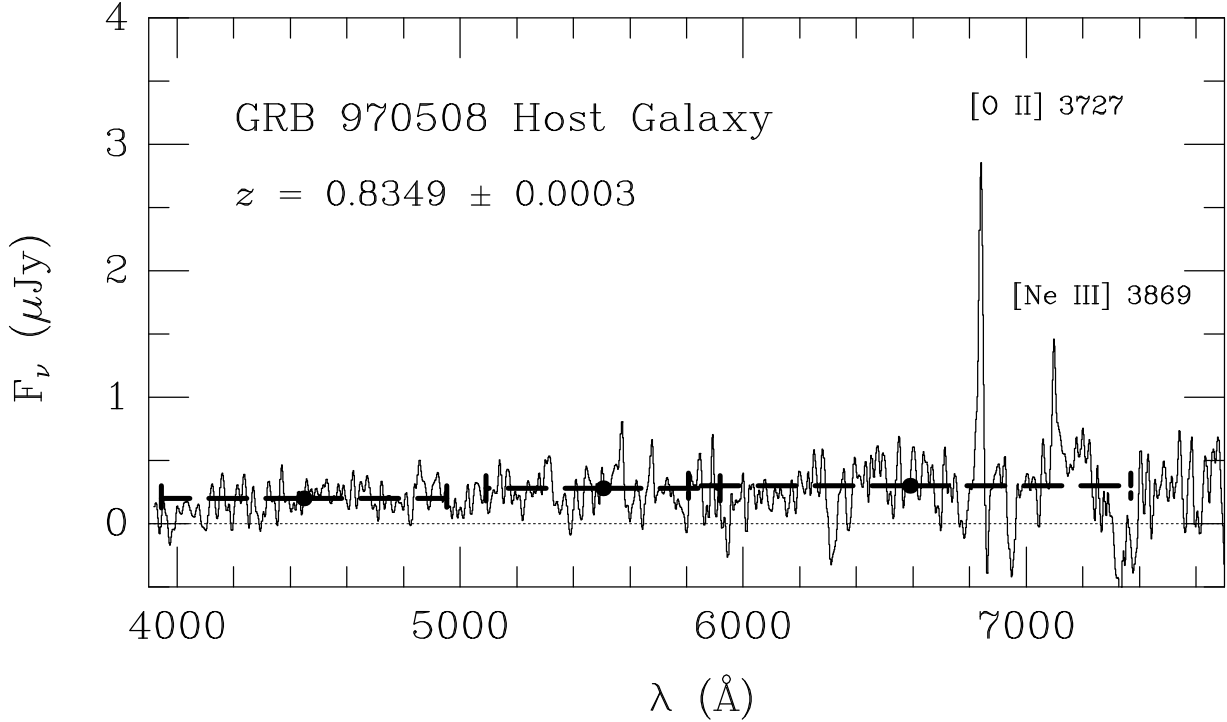}
\caption{A comparison of the GRB~970508 host galaxy $BVR_\mathrm{c}$ photometry
       with the
      spectrum obtained by Bloom et al. (\cite{Bloom98b}).
      The sizes of the points are according to the errors in the $BVR_\mathrm{c}$
      photometry.
      The FWHM of each band is marked by dashed horizontal lines with bars.}
\label{grb970508Keck_SAO}
\end{figure*}

\begin{figure*}[th]
\centering
\includegraphics[width=17cm,bb=55 190 540 455,clip]{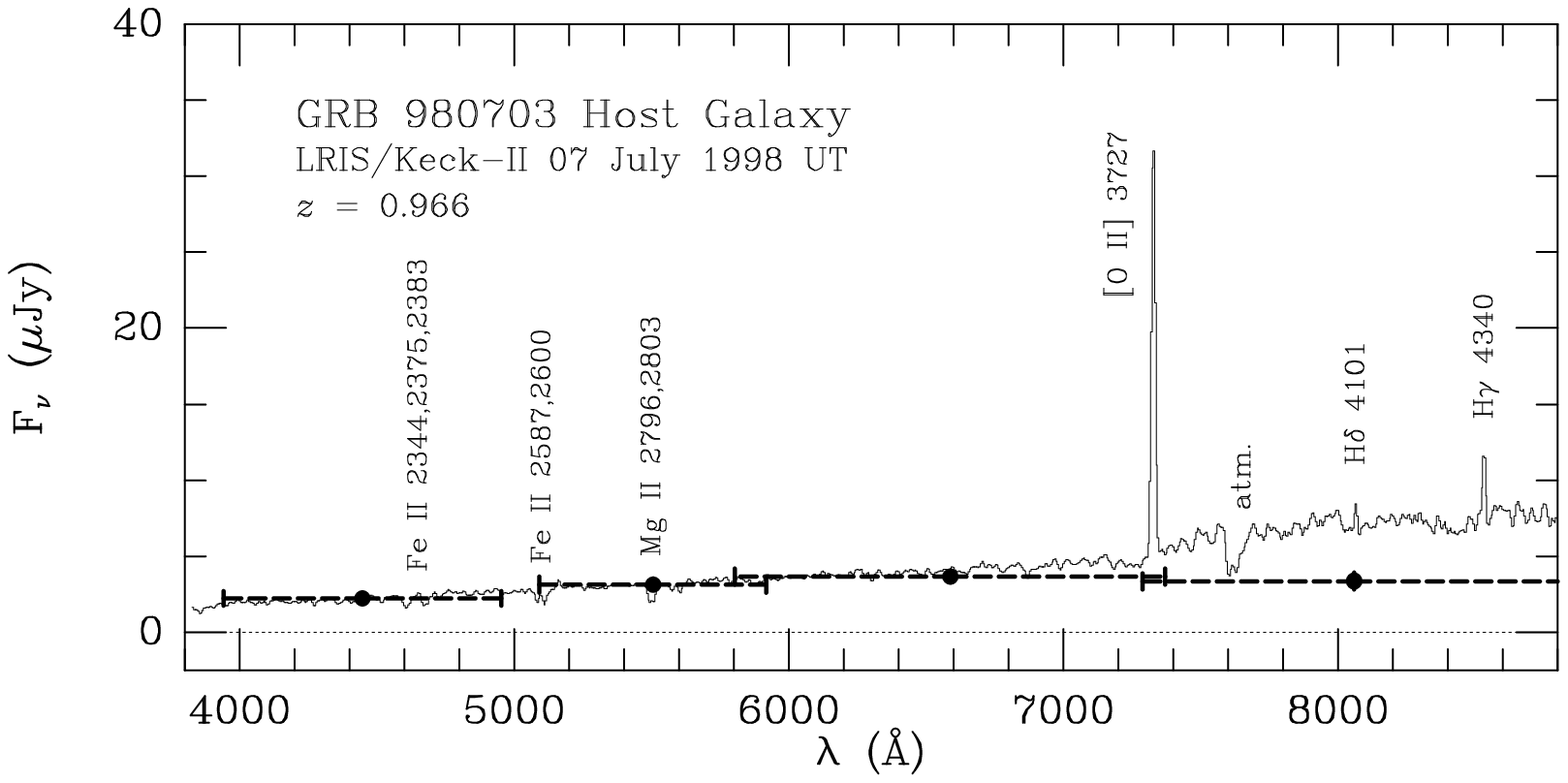}
\\
\includegraphics[width=17cm,bb=55 190 540 450,clip]{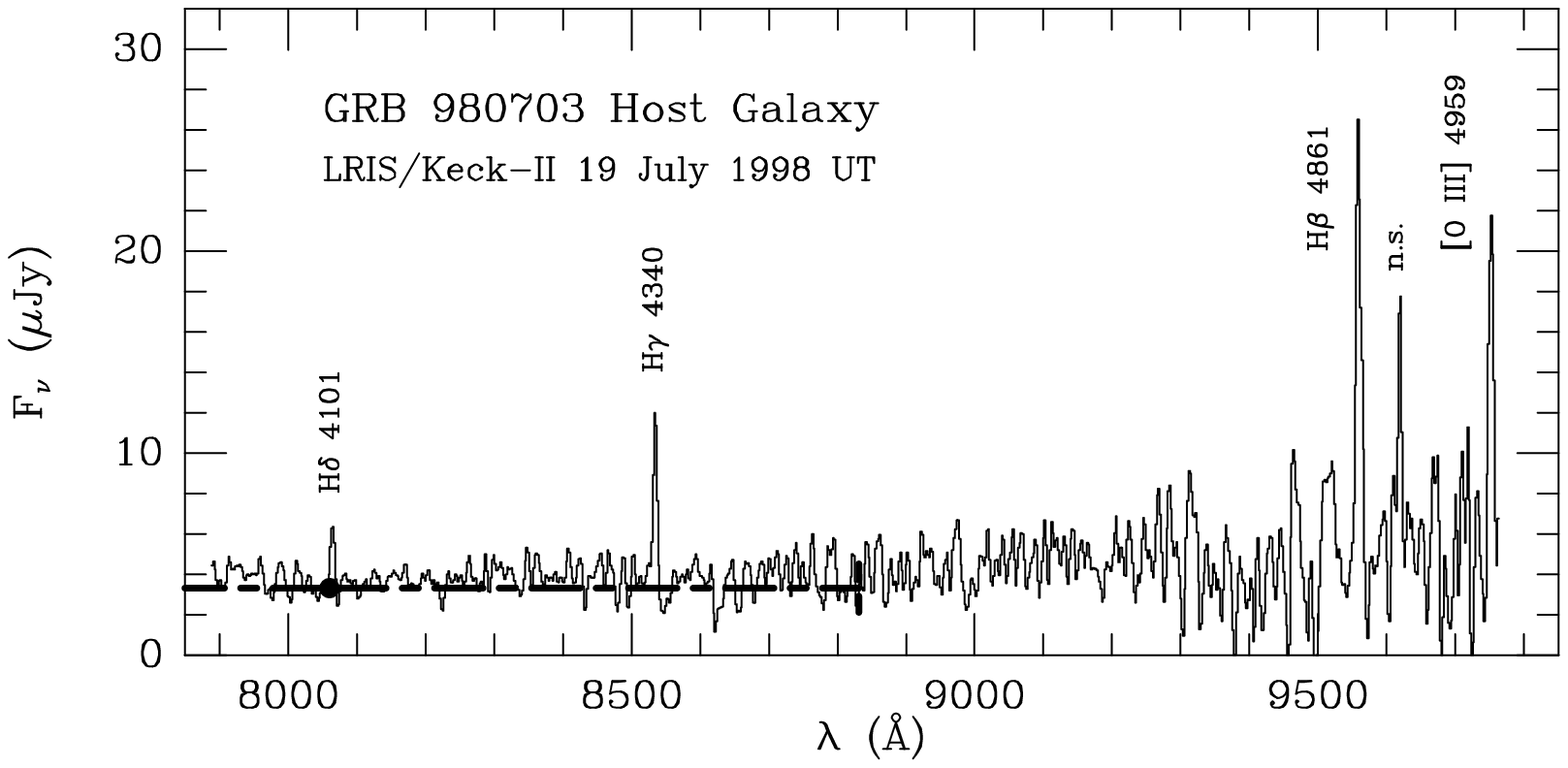}
\caption{ A comparison of the GRB~980703 host galaxy
      $BVR_\mathrm{c}I_\mathrm{c}$ photometry with the
      spectra obtained by Djorgovski et al. (\cite{Djorgovski}).
      The sizes of the points follow the errors in the
      $BVR_\mathrm{c}I_\mathrm{c}$ photometry.
      The FWHM of each band is marked by dashed horizontal lines with bars.
      As we can see from the spectra obtained in {\it two epochs (07 July and
      19 July)}, the central
      intensities of Balmer's lines H$\gamma$ and H$\delta$ remain close
      to constant during the observations in two epochs. This observational
      fact can be explained if the OT shone through an \ion{H}{ii} region
       displaying Balmer's lines in absorption.}
\label{grb980703Keck_SAO}
\end{figure*}

\section{Comparison of the $BVR_\mathrm{c}I_\mathrm{c}$ spectra
with the SEDs of local starburst galaxies}\label{BVRI-SB}
As we noted in Sect.~1, our aim is to determine
the parameters (i.e. the luminosities) of the host galaxies.
To estimate the reddening in the hosts (in {\it a first approximation}),
the SEDs of local starburst galaxies with known colour excess $E(B-V)$ were used.
In addition, the comparison with these SEDs allows us to estimate the
K-correction and, then, the luminosities of the host galaxies (it will be
done in Sect.~\ref{Est-Mabs}). In addition, we compared our
$BVR_\mathrm{c}I_\mathrm{c}$  broadband
flux spectra to the spectra of different Hubble-type galaxies
(from E-SO to Im, see Sokolov et al. \cite{Sokolov1999}; Sokolov et al.
\cite{Sokolov2000}
and found that our data are best fitted by the spectra of starburst galaxies.
By definition, we state that starburst galaxies are galaxies with a strong
UV continuum, which leads to blue observed colours, and with strong nebular
emission lines associated with starburst activity.
Starburst galaxies also undergo a stage of massive star-forming.

To compare our broad-band flux spectra we used the S1, S2, S3, S4, S5, S6
averaged spectral energy
distrubutions (SEDs) for the local starburst galaxies from Calzetti et al.
(\cite{Calz1994}).
The spectra of the local starbursts
were grouped according to increasing values of
the colour excess $E(B-V)$: from S1, with $E(B-V)=0.05$ mag to S6, with
$E(B-V)=0.7$ mag. Using the relation for $\tau_\mathrm{B}^l$
(Balmer optical depth) from
Calzetti et al. (\cite{Calz1994}), we derived the values of colour excess for
an individual group of starburst galaxies. These are: $E(B-V)<0.10$ mag
for S1, $0.11<E(B-V)<0.21$ mag
for S2,  $0.25<E(B-V)<0.35$ mag for S3, $0.39<E(B-V)<0.50$ mag
for S4,
$0.51<E(B-V)<0.60$ mag for S5 and $0.61<E(B-V)<0.70$ mag
for S6 (see Table~3 in Calzetti et al. \cite{Calz1994}).
It should be noted that these SEDs are not observed but are the templates
that have been constructed using real observed starburst SEDs
up to a redshift of $z  \sim 0.03$ (Connoly et al. \cite{Connoly}).

The fluxes of the starburst SEDs were integrated with sensitivity
functions of the $BVR_\mathrm{c}I_\mathrm{c}$ filters (sensitivity
functions have been used
from Bessel \cite{Bessel}) and the derived values were compared to our observed
fluxes. For each SED the $\chi^2$ was calculated as follows:
$$
   \chi^2 = \sum_i \Big(\frac{f_{\mathrm{host},i}-k\cdot f_{\mathrm{template},i}}
	    {\sigma_{f_{\mathrm{host},i}}}\Big)^2
$$
Here $i$ denotes the filters ($BVR_\mathrm{c}I_\mathrm{c}$),
$f_{\mathrm{host},i}$ is
the flux of GRB
host galaxies in the filter $i$, $f_{\mathrm{template},i}$ is the
flux of the template SED integrated with sentivity function of the filter $i$
at effective wavelength of the filter $i$,
$\sigma_{f_{\mathrm{host},i}}$ is the error of the flux of the GRB host galaxy
in the filter $i$, and $k$ is the normalization coefficient.
The results of comparison are given in Table~\ref{fitting}.
In the last column the label ``$\chi^2/d.o.f.$" means the value of $\chi^2$
divided by the degree of freedom.
Figs.~\ref{grb970508_S5}, \ref{grb980703_S2}
and \ref{grb991208_S5} demonstrate the best fitting
(minimum of $\chi^2/d.o.f$) of our broad band spectra by the starburst
template galaxies. However, we should note that in the case of the host
galaxy of GRB~980703, an S2 template galaxy just roughly fitted the
$BVR_\mathrm{c}I_\mathrm{c}$ data.
But, as will be shown in Sect.~\ref{Modelling}, the results of this comparison
will allow
us to derive a more exact fitting of the observational data by modeling.

\begin{table}
\centering
\caption{The results of comparison with local starbursts}
\label{fitting}
\begin{tabular}{lcccc}
\hline
\hline
Host       & template  & $E(B-V)$   & $A_\mathrm{V}^{*}$ & $\chi^2/d.o.f.$ \\
\hline
GRB~970508 & S5        & 0.51--0.60 & 1.5--1.86          & 0.64/3 \\
GRB~980613 & S6        & 0.61--0.70 & 1.8--2.17          & 1.22/2 \\
GRB~980703 & S2        & 0.11--0.21 & 0.3--0.65          & 7.67/3 \\
GRB~990123 & S1        & $<$0.10    & $<$0.31            & 2.29/3 \\
GRB~991208 & S5        & 0.51--0.60 & 1.5--1.86          & 0.99/3 \\
\hline
\hline
\end{tabular}
\parbox[t]{\hsize}{$^{*}$ The values correspond to the Milky Way
extinction law (Cardelli et al. \cite{Card}).}
\end{table}

\begin{figure}
\resizebox{\hsize}{!}{\includegraphics[angle=-90,clip]{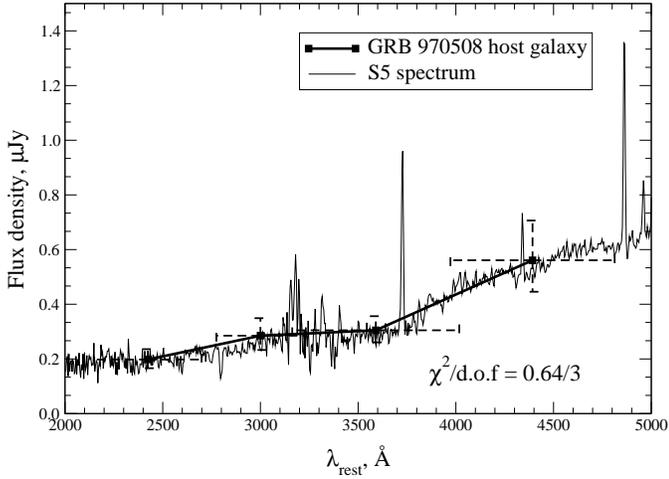}}
\caption{A comparison of the GRB~970508 host galaxy broad-band
      rest-frame ($z=0.835$) flux
      spectrum with the SED of S5 template galaxies
      (see Connoly et al. \cite{Connoly}). The flux of the S5 template was
      scaled to obtain the  best fit.
      Taking into account $z$, the FWHM of each filter for
      $\lambda_\mathrm{eff}$
      is marked by dashed horizontal lines with bars.}
\label{grb970508_S5}
\end{figure}
\begin{figure}
\resizebox{\hsize}{!}{\includegraphics[clip]{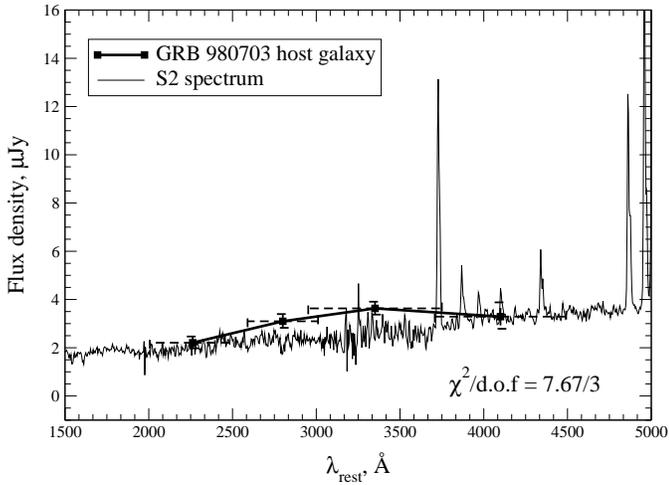}}
\caption{A comparison of the GRB~980703 host galaxy broad-band
      rest-frame ($z=0.966$) flux
      spectrum with the SED of S2-galaxies
      (see Connoly et al. \cite{Connoly}).
}
\label{grb980703_S2}
\end{figure}
\begin{figure}
\resizebox{\hsize}{!}{\includegraphics[width=\hsize,clip]{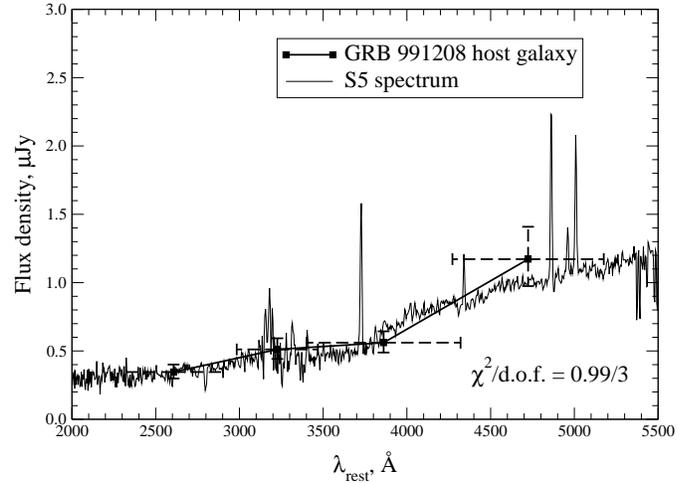}}
\caption{A comparison of the GRB~991208 host galaxy broad-band rest-frame
      ($z=0.7063$) flux
      spectrum with the SED of S5-galaxies
      (see Connoly et al. \cite{Connoly}).
}
\label{grb991208_S5}
\end{figure}
(It should be noted that the spectrum of the host galaxy of GRB~970508 has
no feature at 3200\AA\ like the one seen in S5 (Connoly et al. \cite{Connoly}).
But, we made an estimate of the contribution of this bump for the
S5 template to the $B$ and $V$ bands and concluded that it is negligible).

\section{Estimations of the absolute magnitudes}\label{Est-Mabs}
It is of  great
interest to compare the luminosities of the host galaxies to the
luminosities corresponding to the luminosity of a typical galaxy ($L_*$)
at the ``knee" of the local galaxies luminosity function.
Using the best fit SEDs of
local starbursts from Sect.~\ref{BVRI-SB},
we make estimates of observed luminosities of the 5-th GRB hosts.
We use here mainly a Friedmann model with $\Omega_\mathrm{M}=0.3$ and
$\Omega_\Lambda=0.7$ (see e.g. de Bernardis et al. \cite{Bernardis}).
Recent studies of the Hubble constant put it within the range
$50-70$\,km\,s$^{-1}$\,Mpc$^{-1}$ (see Theureau et al. \cite{Theureau}).
In this paper we adopt $H_0$=60\,km\,s$^{-1}$\,Mpc$^{-1}$.
But to estimate the influence of
the cosmological parameters, we, in addition, use three models
which conveniently limit the possibilities: A (60,1,0),
B (60,0,0) and C (60,0,1), with ($H_0$,$\Omega_\mathrm{M}$,$\Omega_\Lambda$).
For the models the relation
$\Omega_\mathrm{M} + \Omega_\Lambda + \Omega_\mathrm{k} = 1$
is valid, where $\Omega_M = \rho_0 8 \pi G/3H_0^2$, $\Omega_\Lambda=
 \Lambda c^2/3H_0^2$, and $\Omega_\mathrm{k} = -kc^2/R_0^2H_0^2$.  Here $\rho$,
$\Lambda$, $k$, and $R$ are the density, cosmological constant, curvature
constant, and radius of curvature, respectively, and ``0" denotes the present
epoch.

According to our definitions of spectral types of the host galaxies (S1, S2 ...),
we can estimate the K-correction. The definition of the K-correction
(see Oke \& Sandage, \cite{Oke}) is:
\begin{equation}
  K_i = 2.5\ \log\Big\{ (1+z)\frac{\int F(\lambda)S_i(\lambda)d\lambda}
     {\int F(\lambda/(1+z))S_i(\lambda)d\lambda}\Big\}\mbox{,}
\label{k-corr}
\end{equation}
where $F(\lambda)$ is the rest-frame SED,
$S_i(\lambda)$ is the transmission curve for filter $i$ and
$z$ is the redshift.
Using the SEDs from Calzetti et al. (\cite{Calz1994}), we estimated
the $K$-correction for the $B$ band. The results are given in Table~\ref{abs-magn}.
These estimates allow us to derive the absolute magnitudes of
the host galaxies. Using the $B$ magnitudes from Table~\ref{phot} and the
K-correction given above, we derive the
absolute magnitudes given in Table~\ref{abs-magn}. 
We present the above-mentioned absolute magnitudes
for the (A), (B) and (C) cosmological models. They reflect the uncertainty in the 
definition of $M_{\mathrm{B}_\mathrm{rest}}$
due to the variation of the cosmological parameters.
\begin{table*}
\caption{The $K$-corrections and absolute magnitudes of the 6-th GRB host galaxies}
\label{abs-magn}
\centering
\begin{tabular}{clcccc}
\hline
\hline
Host        & $K_\mathrm{B}$-correction   &  \multicolumn{4}{c}{Absolute magnitudes}  \\
\,          &                    & $M_{\mathrm{B}_\mathrm{rest}}$ & (A)    &   (B)  &   (C)  \\
\hline
GRB~970508  &   $0.44$           &   $-18.62$     &$-18.08$&$-18.53$&$-19.09$\\
	    &   $0.51\pm0.32$    &   $-18.69$     &        &        &        \\
	    &$0.0$ ($I_\mathrm{c}$ to $B$)&   $-19.88$     &        &        &        \\
GRB~980613  &   $0.85$           &   $-20.76$     &$-20.14$&$-20.72$&$-21.38$\\
GRB~980703  &   $-0.01$          &   $-21.27$     &$-20.60$&$-21.12$&$-21.73$\\
GRB~990123  &   $0.13$           &   $-20.93$     &$-20.20$&$-21.02$&$-21.82$\\
GRB~991208  &   $0.46$           &   $-18.79$     &$-18.29$&$-18.68$&$-19.18$\\
	    &   $0.21\pm0.31$    &   $-18.54$     &        &        &        \\
	    &$0.0$ ($I_\mathrm{c}$ to $B$)&   $-19.81$     &        &        &        \\
\hline
\hline
\end{tabular}
\end{table*}

For the host galaxies of GRB~970508 ($z=0.835$) and GRB~991208
(z=0.7063), we give two additional estimates of $K_\mathrm{B}$.
As  the $I_\mathrm{c}$ band roughly corresponds to the
$B$ band in the rest frame, we can calculate directly from Eq.
(\ref{k-corr}) a second estimate of the K-correction for the $B$-magnitude, 
assuming
$2.5\ \log(F_{\mathrm{B}_\mathrm{rest}}/F_{\lambda_{\mathrm{B}/(1+z)}})
\approx 2.5\ \log(F_{\mathrm{I}_\mathrm{obs}} /F_{\mathrm{B}_\mathrm{obs}})$.
Thus we derived:  $K_\mathrm{B} = 0.51\pm 0.32$ mag and
$K_\mathrm{B} = 0.21\pm 0.31$ mag
for the host galaxies of GRB~970508 and GRB~991208, respectively.
As a third estimate we can assume  $K_\mathrm{B}$= $0.0$ mag
considering that
the observed $I_\mathrm{c}$ filter matchs the $B$ filter in the rest-frame.
This differs from the
other two estimates, since going from $I_\mathrm{c}$ to the $B$-band we have
to take into account the mismatch of zero points and the transmission curves
of the colour bands.

Table~\ref{abs-magn} displays the uncertainties in
the luminosities (by about one mag) due to different cosmological models. 
Therefore, the final values of the absolute magnitudes of the GRB host galaxies
are given in Table~\ref{summary} for a more realistic
Friedmann model with $H_0$=60\,km\,s$^{-1}$\,Mpc$^{-1}$,
$\Omega_\mathrm{M}$=0.3 and $\Omega_\Lambda$=0.7.
The derived luminosities allow us to perform the theoretical modeling
of continuum SEDs of the host galaxies.

We notice that in Sect.~\ref{BVRI-SB}
the same best $BVR_\mathrm{c}I_\mathrm{c}$-fits were obtained by the template
SEDs for the host galaxies of GRB~970508 and GRB~991208.
It means that these two galaxies might have a similar internal extinction
(see Table~\ref{fitting} and Sect.~\ref{Est_SFR}).
Moreover, as can be seen from Table~\ref{abs-magn}, these two galaxies have
similar absolute magnitudes.

\section{Modeling of the SEDs for GRB hosts}\label{Modelling}
To estimate the effects of the internal
extinction, we performed a theoretical modeling of the continuum spectral energy
distributions of the host galaxies of GRB~970508 and GRB~980703.
Note that for these galaxies the spectra, $BVR_\mathrm{c}I_\mathrm{c}$ and
near-infrared data are
available and, what is most essential before drawing
any conclusions, these galaxies have the lowest (GRB~970508) and
the highest (GRB~980703) luminosities (see Table~\ref{abs-magn}).
Using a population synthesis modeling, we constructed a set of model
theoretical templates for these two host galaxies
as {\it a second approximation } to estimate internal extinction.

\subsection{The method}\label{method}
We used the PEGASE code
({\bf P}rojet d'{\bf \'E}tude des {\bf GA}laxies
par {\bf S}ynth\`ese {\bf \'E}volutive)
developed by Fioc \& Rocca-Volmerange (\cite{PEGASE}),
which is publicly available at
\verb"ftp://ftp.iap.fr"
\verb"/pub/from_users/fioc/PEGASE/PEGASE.2/").
We assumed the Salpeter IMF (Initial Mass Function) with the low and high
mass cut-offs to be 0.1\,\Msun and 120\,\Msun, respectively.
In the computations the $Z_\odot$ and $0.1Z_\odot$ metallicity were assumed
as well as the simplest instantaneous burst and more complex exponential
decreasing star formation.
Since PEGASE computes SEDs at z=0, i.e. presents data in monochromatic
luminosities, we used
the cosmological parameters $H_0$=60\,km\,s$^{-1}$\,Mpc$^{-1}$,
$\Omega_\mathrm{M}$=0.3 and $\Omega_\Lambda$=0.7 for trasforming
monochromatic luminosities into observed fluxes.

The comparison of the $BVR_\mathrm{c}I_\mathrm{c}$ broad-band flux
spectra of the host
galaxies of GRB~970508, GRB~980703, GRB~990123 and GRB~991208 with the
local starburst SEDs allows us to consider that
there is significant internal extinction in the host galaxies. To estimate
the influence of this important fact we used the Cardelli et al. (\cite{Card})
and Calzetti et al. (\cite{Calz2000}) extinction laws. These laws differ in
the shape of the curve and the parameterization of
$R_\mathrm{V}\equiv A_\mathrm{V}/E(B-V)$ with
$R_\mathrm{V}=4.05$ for the Calzetti et al. law and $R_\mathrm{V}=3.1$
for the Cardelli
et al. law. Fig.~\ref{extin_laws} shows both extinction law curves.
Notice that the Cardelli et al. extinction law represents the reddening in
the Milky Way (MW),
while the Calzetti et al. law was derived empirically from a sample of
integrated spectra of starburst galaxies (SB).
\begin{figure}
\resizebox{\hsize}{!}{\includegraphics[clip]{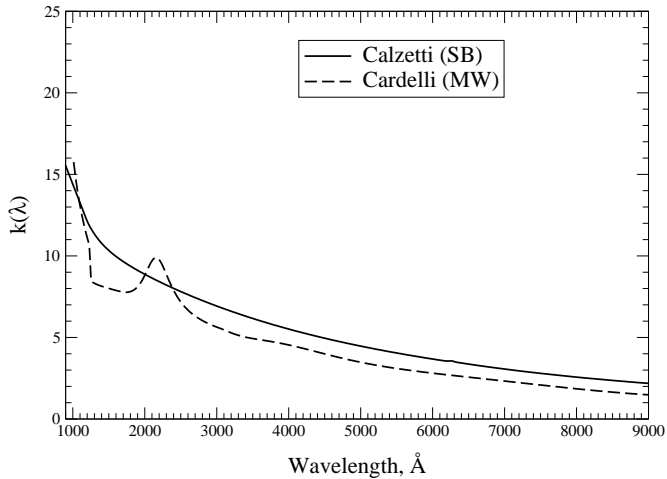}}
\caption{Extinction curves $k(\lambda)$ for the Cardelli et al. (\cite{Card})
and Calzetti et al. (\cite{Calz2000}) reddening laws.
Cardelli et al. extinction law represents the reddening in Milky Way,
while Calzetti et al. law was derived empiracally from a sample of
integrated spectra of starburst galaxies.
}
\label{extin_laws}
\end{figure}
For simplicity we assumed a dust-screen model with the following
mathematical relation for the reddening of spectra:
$F_\mathrm{obs}(\lambda) = F_\mathrm{int}(\lambda)10^{-0.4A_\lambda}$,
where $F_\mathrm{obs}$ and $F_\mathrm{int}$ are the observed and intrisic
fluxes, respectively, and
$A_\lambda = k(\lambda)E(B-V)$ is the extinction at a wavelength $\lambda$.

For calculation of the theoretical template SEDs we applied a
two-component model:
the first component is just a burst (``burst" component) of star-formation
and the second one is an old (``old" component) stellar population.
Both give corresponding contributions to a continuous
spectrum, which produces on the aggregate the observed energy distribution
in the galaxy spectrum.
The ``burst" component responds to emission lines and a nebular continuum.
For this reason, we roughly fixed the ``burst" component parameters using
the luminosity of \ion{O}{ii} forbidden emission lines by fitting to the
observed fluxes
and taking into account the assumed reddening laws. In the modeling, we
used the observed fluxes of the \ion{O}{ii} emmision line from
Bloom et al. (\cite{Bloom98b}) and Djorgovski et al. (\cite{Djorgovski}),
which are $(2.98\pm0.22)\cdot10^{-17}$\,erg\,s$^{-1}$\,cm$^{-2}$ and
$(30.4\pm3)\cdot10^{-17}$\,erg\,s$^{-1}$\,cm$^{-2}$ for the host galaxy
of GRB~970508 and GRB~980703, respectively.
In addition, we estimated the contribution of the emission lines
(not only [\ion{O}{ii}]) of the
spectra of the host galaxy of GRB~970508 and GRB~980703 to broad-band
magnitudes and found that it is negligible for our
$BVR_\mathrm{c}I_\mathrm{c}$ errors.
The choice of only forbidden
lines allows us to avoid effects of contribution of stellar absorption into
luminosity of emission lines, for example in the cases of Balmer lines.
To see that only the short-duration star-formation burst (or ``burst" component)
determines the emission spectrum of the galaxy,
it would be interesting to trace spectral evolution of a burst
of star-formation with time.
According to PEGASE, the nebular emission (lines and continuum) of an
{\it instantaneous burst}, e.g.,  dominates up to $\approx$~10\,Myr
(Fig.~\ref{evol_z0.02}). The fixed parameters of a star-formation burst
allow us to determine the age and mass of the oldest stellar population
(``old" component) by fitting to the observed
$BVR_\mathrm{c}I_\mathrm{c}$ fluxes.
Of course, we bear in mind the contributions
into  $BVR_\mathrm{c}I_\mathrm{c}$ of the
``burst" component. We reddened the burst light alone (in the ``burst"
component for the {\it instantaneous burst} as well as for the
{\it exponential decreasing} star formation scenarios),
since the extinction in starburst regions is higher
than the average extinction and the colour excess $E(B-V)$ is derived
from emission of Balmer lines
(Calzetti et al. \cite{Calz1994}) associated with \ion{H}{ii} regions.

\begin{figure}
\resizebox{\hsize}{!}{\includegraphics[angle=-90,bb=50 50 554 770,clip]{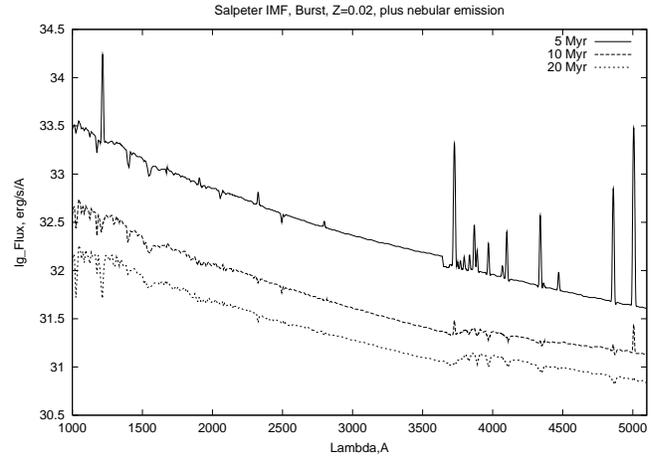}}
\caption{Spectral evolution of an instantaneous burst of 1\,\Msun
for 5, 10 and 15 {\bf Myr} after the event. Nebular emission is consider here.
The model monochromatic flux is in units of erg/s/\AA/\Msun
(see Fioc \& Rocca-Volmerange \cite{PEGASE}).}
\label{evol_z0.02}
\end{figure}

\subsection{The host galaxy of GRB~970508}\label{Modeling_GRB970508}
As was shown in Sect.~\ref{BVRI-SB}, the GRB~970508
$BVR_\mathrm{c}I_\mathrm{c}$ broad-band flux
spectrum is well approximated by the S5 average SEDs
and our photometric points are in good agreement with the continuum of the
host galaxy shown in Fig.~\ref{grb970508Keck_SAO}. Thus, we have to
determine the colour excess $E(B-V)$.
So, the definition of the host galaxy as an S5 averaged starburst group galaxy
allows us to estimate the value of $A_\mathrm{V}$. According to
$E(B-V)=0.5-0.60$ mag
we derive $A_\mathrm{V}=2.0-2.43$ mag for
Calzetti et al. (\cite{Calz2000}) and
$A_\mathrm{V}=1.5-1.86$ mag for Cardelli et al. (\cite{Card})
reddening laws, respectively.

First, we performed the theoretical modeling assuming the Calzetti et al.
extinction law.
We used both the {\it instantaneous burst} and the {\it exponential decreasing}
star formation
scenarios as well as both $Z_\odot$ and $0.1Z_\odot$ metallicities.
In Figs.~\ref{GRB970508_Model_Calz}, \ref{GRB970508_Model_Calz_best}
the results of modeling for the Calzetti et al. reddening law are presented.
To constrain our models, we plot in these Figures, in addition,
the upper limit of HST/NICMOS $H$-band flux from
Pian et al. (\cite{Pian}). The parameters of the resulting theoretical
template SEDs corresponding to the minimum of $\chi^2$ are given in
Table~\ref{model_GRB970508_Calz}. In
Table~\ref{model_GRB970508_Calz}
the first column indicates the scenario of star formation,
in the second and third columns the age and mass of the old component (see text)
are given and in the fourth and fifth columns the same parameters of the burst
component are presented.
For the exponential decreasing star formation scenario
we also present the values of $\tau$, which is the time when the star formation
rate decreases by the factor of {\it e}.
The sixth column denotes the metallicity of the theoretical
template, the seventh column contains the model value of the
[\ion{O}{ii}]3727\AA\,
emission line flux. We present the minimal value of $\chi^2$ in the
eighth column and in the last column the best fit values of extinction
$A_\mathrm{V}$ are given.

As was described in \ref{method}, we performed the theoretical modeling
making use of Cardelli et al. extinction law.
We again assumed both star formation scenarios and $Z_\odot$ and
$0.1Z_\odot$ metallicities. The resulting template SEDs with minimal $\chi^2$
are presented in Fig.~\ref{GRB970508_Model_Card},
\ref{GRB970508_Model_Card_best}.
The parameters of these SEDs are summarized
in Table~\ref{model_GRB970508_Card}.
The organization of the Table is the
same as that of Table~\ref{model_GRB970508_Calz}.

It is important to note that we checked the values of $\chi^2$ out
of the range of $A_\mathrm{V}$ given by fitting to S5-template and found
that the best fit theoretical templates correspod to the values of
$A_\mathrm{V}$, which lie in the S5-range.
The results of the modeling imply that there is some uncertainty in the
choice of the reddening laws. At least, as can be seen from
Fig.~\ref{GRB970508_Model_Calz}, we may say that only one theoretical
template with a metallicity of $0.1Z_\odot$ and exponential decreasing scenario
for Calzetti et al. reddening law
is not consistent with the upper limit of the $H$-band and, thus, can be ruled
out. A more detailed analysis of the results of the modeling is given in
Sect.~\ref{Disc}.

\begin{figure}
\resizebox{\hsize}{!}{\includegraphics[clip]{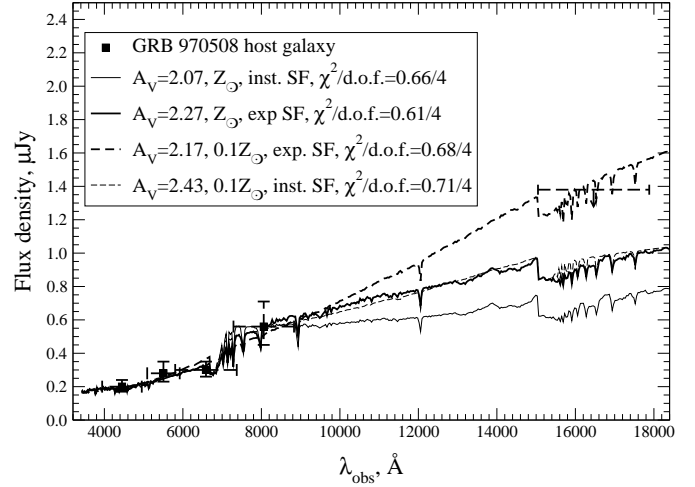}}
\caption{
The best fit for the SED model to the $BVR_\mathrm{c}I_\mathrm{c}$
photometry of the GRB 970508 host galaxy,
assuming the Calzetti et al. (\cite{Calz2000}) extinction law. Also the
upper limit of HST/NICMOS $H$-band is plotted (see Pian et al. \cite{Pian}).
The observed wavelengths are given.}
\label{GRB970508_Model_Calz}
\end{figure}
\begin{figure}
\resizebox{\hsize}{!}{\includegraphics[bb=30 25 730 540,clip]{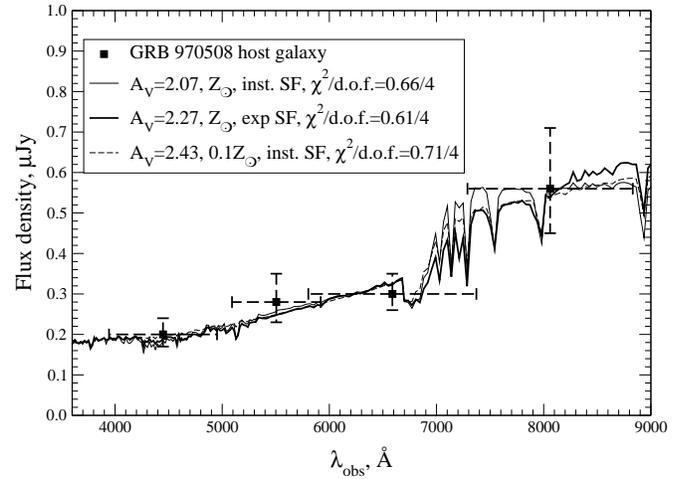}}
\caption{
The best fit model SED assuming the Calzetti et al. extinction law.
The observed wavelengths are given.
}
\label{GRB970508_Model_Calz_best}
\end{figure}

\begin{figure}
\resizebox{\hsize}{!}{\includegraphics[clip]{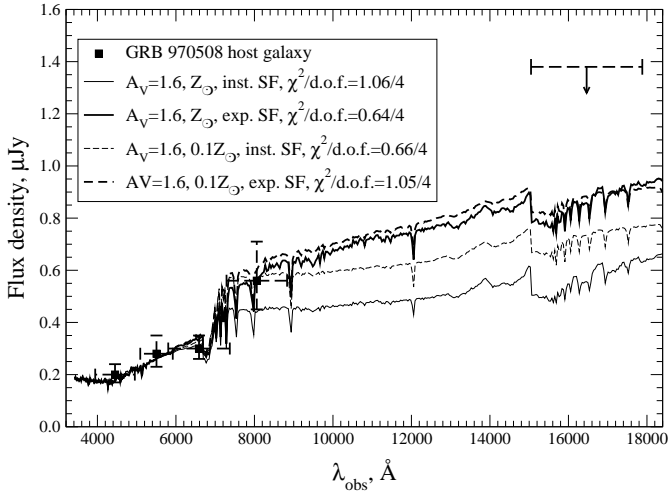}}
\caption{
The best fit for the SED model to the $BVR_\mathrm{c}I_\mathrm{c}$
photometry of the GRB 970508 host galaxy,
assuming the Cardelli et al. (\cite{Card}) extinction law. Also the
upper limit of HST/NICMOS $H$-band is given (see Pian et al. \cite{Pian}).
The observed wavelengths are given.}
\label{GRB970508_Model_Card}
\end{figure}
\begin{figure}
\resizebox{\hsize}{!}{\includegraphics[bb=30 25 730 540,clip]{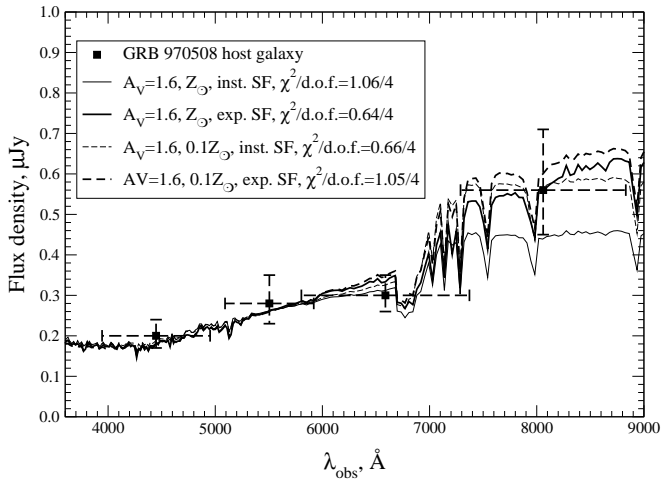}}
\caption{
The best fit model SED assuming Cardelli et al. extinction law.
The observed wavelengths are given.}
\label{GRB970508_Model_Card_best}
\end{figure}

\begin{table*}
\centering
\caption{The parameters of the theoretical template SEDs corresponding to
minimum of $\chi^2$ for the host galaxy of GRB~970508 assuming Calzetti et al.
extinction law$^*$}
\label{model_GRB970508_Calz}
\begin{tabular}{lllllrlll}
\hline
\hline
Scenario      & \multicolumn{2}{l}{old component} & \multicolumn{2}{l}{burst component}    & metallicity& [\ion{O}{ii}] model & $\chi^2_\mathrm{min}$/d.o.f. & $A_\mathrm{V}$ \\
	      &  Age, Gyr       &  Mass, \Msun~~~  &  Age, Myr        &  Mass, \Msun       &            & flux$^{**}$         &                       &       \\
\hline
Instantaneous &   0.14          &$3.16\cdot10^9$   &  1               & $5.62\cdot10^7$    & $Z_\odot$  & $2.98\cdot10^{-17}$ & 0.66/4                & 2.07  \\
burst         &   0.08          &$1.00\cdot10^8$   &  4               & $1.78\cdot10^8$    &$0.1Z_\odot$& $2.92\cdot10^{-17}$ & 0.71/4                & 2.43  \\
	      &                 &                  &                  &                    &            &                     &                       &       \\
Exponential   &1.2 ($\tau=500$Myr)&$1.00\cdot10^9$ & 3 ($\tau=300$Myr)& $5.62\cdot10^9$    & $Z_\odot$  & $2.61\cdot10^{-17}$ &0.61/4                 & 2.27  \\
decreasing    &0 ($\tau=400$Myr)&$1.00\cdot10^9$   &80 ($\tau=400$Myr)& $5.62\cdot10^9$    &$0.1Z_\odot$& $2.96\cdot10^{-17}$ &0.68/4                 & 2.17  \\

\hline
\hline
\end{tabular}
\parbox[t]{\textwidth}{
$^*$ The parameters were obtained by fitting to
$BVR_\mathrm{c}I_\mathrm{c}$ data with observed
flux of [\ion{O}{ii}] emission line. \\
$^{**}$ The flux of the [\ion{O}{ii}] emission line derived in modeling (in units
of erg\,s$^{-1}$\,cm$^{-2}$).
}
\end{table*}

\begin{table*}
\centering
\caption{The parameters of the theoretical template SEDs corresponding to
minimum of $\chi^2$ for the host galaxy of GRB~970508 assuming Cardelli et al.
extinction law}
\label{model_GRB970508_Card}
\begin{tabular}{lllllrlll}
\hline
\hline
Scenario      & \multicolumn{2}{l}{old component} & \multicolumn{2}{l}{burst component}    & metallicity& [\ion{O}{ii}] model & $\chi^2_\mathrm{min}$/d.o.f. & $A_\mathrm{V}$ \\
	      &  Age, Gyr       &  Mass, \Msun~~~  &  Age, Myr        &  Mass, \Msun       &            & flux$^*$            &                       &       \\
\hline
Instantaneous &   0.08          &$1.78\cdot10^8$   &  0               & $3.16\cdot10^7$    & $Z_\odot$  & $3.01\cdot10^{-17}$ & 1.06/4                & 1.6   \\
burst         &   0.16          &$3.16\cdot10^8$   &  0               & $3.16\cdot10^7$    &$0.1Z_\odot$& $3.00\cdot10^{-17}$ & 0.66/4                & 1.6   \\
	      &                 &                  &                  &                    &            &                     &                       &       \\
Exponential   &1.0 ($\tau=400$Myr)&$1.00\cdot10^9$ & 0 ($\tau= 10$Myr)& $3.16\cdot10^8$    & $Z_\odot$  & $3.01\cdot10^{-17}$ & 0.64/4                & 1.6   \\
decreasing    &1.2 ($\tau=450$Myr)&$1.00\cdot10^9$ & 0 ($\tau= 10$Myr)& $3.16\cdot10^8$    &$0.1Z_\odot$& $3.00\cdot10^{-17}$ & 1.05/4                & 1.6   \\

\hline
\hline
\end{tabular}
\parbox[t]{\textwidth}{
$^*$ The flux of the [\ion{O}{ii}] emission line derived in modeling (in units
of erg\,s$^{-1}$\,cm$^{-2}$).
}
\end{table*}

\subsection{The host galaxy of GRB~980703}\label{Modeling_GRB980703}
The comparison of our $BVR_\mathrm{c}I_\mathrm{c}$ broad-band flux
spectrum with the spectra
obtained in two epochs (see Fig.~\ref{grb980703Keck_SAO}) has shown that
on 7 July the contribution of the OT near the [\ion{O}{ii}] emission line was
not considerable and we may conclude that the luminosity of the [\ion{O}{ii}]
line is not
affected by the OT. Moreover, from the
19 July data, the spectra of the host galaxy are already not contaminated by
light from the OT. Thus, in spite of the fact that our observations were
carried out $\sim$ 20 days after the GRB event, we may consider that our
$BVR_\mathrm{c}I_\mathrm{c}$
broad-band flux spectrum is exactly that of the host galaxy without any
contribution from the OT.

As we determinated above, the host galaxy is an S2 average
local starburst with $E(B-V)=0.1-0.21$ mag. According to the $E(B-V)$ we have
computed $A_\mathrm{V}=0.4-0.85$ mag for the Calzetti et al.
(\cite{Calz2000}) reddening law and
$A_\mathrm{V}=0.3-0.65$ mag for the Cardelli et al. (\cite{Card}) one,
which is in agreement with the value of 0.3$\pm$0.3 mag. obtained by
Djorgovski et al. (\cite{Djorgovski}) from the ratios of the H lines.

We performed the theoretical modeling using at first the simplest
{\it instantaneous burst} of star formation scenario for the Sun metallicity.
The best fit parameters (by the $BVR_\mathrm{c}I_\mathrm{c}$ fluxes) were
derived as follows:
for the ``burst" component of the model the
age is 3.0\,Myr and the mass is $1.78\cdot10^8$\,\Msun,
and for the ``old" component (for the old stellar
population) the age is 2.5\,Gyr and the mass is $1.78\cdot10^{10}$\,\Msun.
Fig.~\ref{GRB980703_Model_Card} shows the results of the modeling.
It should be noticed that our $BVR_\mathrm{c}I_\mathrm{c}$ data are
better fitted by the theoretical spectrum rather than the
S2-template starburst galaxy (see Sect.~\ref{BVRI-SB}).
Calculations have shown that the $BVR_\mathrm{c}I_\mathrm{c}$
fluxes are not described by a model using the Calzetti et al. reddening law,
but best fitted by the Cardelli et al.  extinction.
At z=0.966 the effective wavelength of filter $B$ ($\lambda=4448$\AA)
in the rest frame ($\lambda=2260$\AA) lies on
the bump of Cardelli's extinction curve ($\lambda\approx2200$\AA, see
Fig.~\ref{extin_laws}), which allows us to choose
the reddening curve corresponding to the law of Cardelli (MW).
In Fig.~\ref{GRB980703_MODELS_best} it corresponds to the deficit of observed
flux in the $B$-band.

In addition, we performed theoretical modeling
making use of the avalaible infrared data.
Bloom et al. (\cite{Bloom98a}) reported the $JHK$ photometry of the OT
of GRB~980703.
The observations were carried out with the NIRC instrument at the Keck 10-m
telescope on 7 Aug. 1998, e.g., later than our
$BVR_\mathrm{c}I_\mathrm{c}$ observations. Thus, we
may assume that these $JHK$ magnitudes are magnitudes of the host galaxy
with a negligible contribution of the OT.
In Fig.~\ref{GRB980703_BVRIJHK_Model_Card} we plot the SED model
taking into account the infrared observations. Here we used the
zero points of the $JHK$ band from Bessel \& Brett (\cite{Bessel_Brett}).
The parameters of the resulting SEDs are presented in Table~\ref{model_GRB980703}.
For comparison in Fig.~\ref{GRB980703_Model_Card} we present the results of
the modeling using only $BVR_\mathrm{c}I_\mathrm{c}$ data (the parameters
are given above).
As can be seen from the comparison of model results for
$BVR_\mathrm{c}I_\mathrm{c}$ and
$BVR_\mathrm{c}I_\mathrm{c}JHK$
fitting (see Fig.~\ref{GRB980703_BVRIJHK_Model_Card}),
we should use the infrared where available in the modeling for more
exact estimates of mass and age of the ``old" component. Indeed, $JHK$ bands
($\lambda_\mathrm{eff}=12370,16464,22105\,\mbox{\AA\AA}$, respectively)
correspond to the optical window and near infrared in the rest frame
(for $z=0.835$, $\lambda_\mathrm{eff}/(1+z)=6741,8972,12046\,\mbox{\AA\AA}$
for $JHK$, respectively) where the
old stellar population dominates and the contribution of the ``burst" component
(or young stellar population) is small.
The maximum of the burst contribution lies in the UV part of the spectrum which
corresponds to observed $BVR_\mathrm{c}I_\mathrm{c}$ bands. In fact, from
Figs.~\ref{GRB980703_Model_Card} and
\ref{GRB980703_BVRIJHK_Model_Card} we see that the modeling
did not change the UV part of the SED.
Consequently, the estimates of SFR are not affected
in the case where we take into account the infrared data.
In Table~\ref{model_GRB980703} the final results of the modeling taking into
account the $JHK$ photometry of the host galaxy of GRB~980703 are given.

\begin{figure}
\resizebox{\hsize}{!}{\includegraphics[clip]{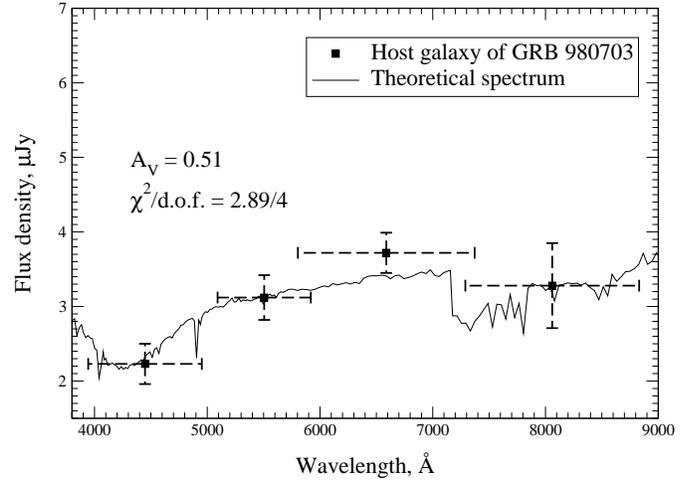}}
\caption{The SED model fitting of only $BVR_\mathrm{c}I_\mathrm{c}$
photometry (without $JHK$)
of the GRB~980703 host galaxy assuming the
Cardelli et al. (\cite{Card}) extinction law.
The observed wavelengths are given.}
\label{GRB980703_Model_Card}
\end{figure}
\begin{figure}
\resizebox{\hsize}{!}{\includegraphics[clip]{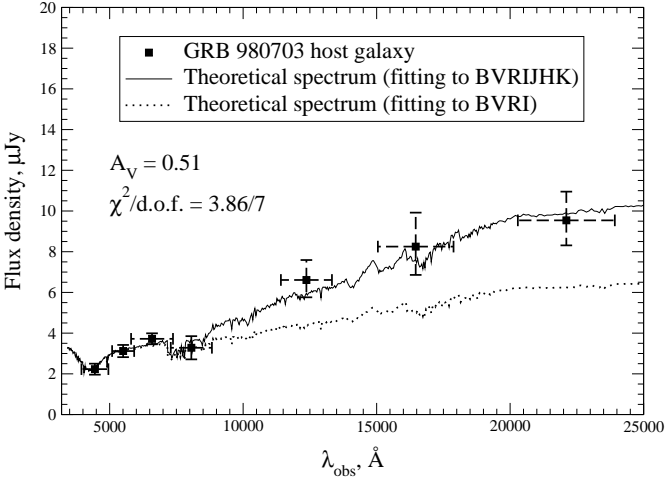}}
\caption{The SED model fitting the $BVR_\mathrm{c}I_\mathrm{c}$ and $JHK$
(Bloom et al. \cite{Bloom98a})
photometry of the GRB~980703 host galaxy assuming the
Cardelli et al. (\cite{Card}) extinction law. Dotted line is the SED model from
Fig.~\ref{GRB980703_Model_Card} (without $JHK$ bands).
The observed wavelengths are given.}
\label{GRB980703_BVRIJHK_Model_Card}
\end{figure}

Following our method we performed the theoretical modeling using
the simplest instantaneous burst as well as more complex {\it exponential
decreasing} star formation scenarios, both ($Z_\odot$ and $0.1Z_\odot$)
metallicities. The results of modeling are presented
in Fig.~\ref{GRB980703_MODELS} and the parameters of the template SEDs are
given in Table~\ref{model_GRB980703}. Here we used
$BVR_\mathrm{c}I_\mathrm{c}$ and $JHK$ data
(with the observed flux in the [\ion{O}{ii}]3727\AA\ emission line).
In Fig.~\ref{GRB980703_MODELS_best} we plot the
$BVR_\mathrm{c}I_\mathrm{c}$ part of the spectrum
of the best fit SEDs.

Table~\ref{model_GRB980703} gives all scenario versions of computations of
template SEDs corresponding to minimum $\chi^2$.
The organization of the Table is the
same as that of Table~\ref{model_GRB970508_Calz} and
Table~\ref{model_GRB970508_Card}.
Fig.~\ref{GRB980703_MODELS}  gives these template SEDs.
As well as in the case of the modeling of the host galaxy of GRB~970508, we
modeled SEDs out of the range of $A_\mathrm{V}$, which corresponds to the
S2 template, and again found that the best fit template
$A_\mathrm{V}$ lies in the
S2-range. As is seen from Fig.~\ref{GRB980703_MODELS}, regardless
of the scenario choice, all template SEDs
with minimum $\chi^2$ correspond to similar values of $A_\mathrm{V}$.
Nevertheless, the best fits are obtained with SEDs of solar metallicity
for the exponential and instantaneous burst scenarios with
$A_\mathrm{V}$=0.51 and 0.64
mag.
The least deviations from observational $BVR_\mathrm{c}I_\mathrm{c}JHK$
fluxes (plus flux in
the line [\ion{O}{ii}]3727\AA) is obtained with SED corresponding to a galaxy
of age of several Gyr. Thus, we choose this variant of SED with
$A_\mathrm{V}=0.64$ mag for calculation of SFR.

But it turns out that the choice of a definite law of reddening is more
essential for the choice of $A_\mathrm{V}$ than the choice of burst scenario,
at least for the case of this GRB~980703 host galaxy.
As was noted above, in the case of the host galaxy of GRB~980703 we have
an opportunity to choose the internal extinction law using the
observational data. Indeed, as is seen from
Fig.~\ref{GRB980703_MODELS_best}, at the $B$-band wavelength
the flux deficit is observed. Using the obtained value of absorption
(see Table~\ref{model_GRB980703}) we can estimate the flux deficit in the
$B$-band
in comparison to the $I$-band as if it was a continuous reddening law
like Calzetti (SB). This $B$-band flux deficit is equal to
40\% -- 50\% for $A_\mathrm{V}=0.51-0.64$ mag, which is really observed
for our $BVR_\mathrm{c}I_\mathrm{c}$
broad-band flux spectrum (see Fig.~\ref{GRB980703_MODELS_best}).

The parameters of the GRB hosts with the account of our analysis for
these two galaxies are given in Table~\ref{final_model} and will be disscused
in more detail in Sect.~\ref{Disc}.

\begin{figure*}
\centering
\includegraphics[width=17cm,clip]{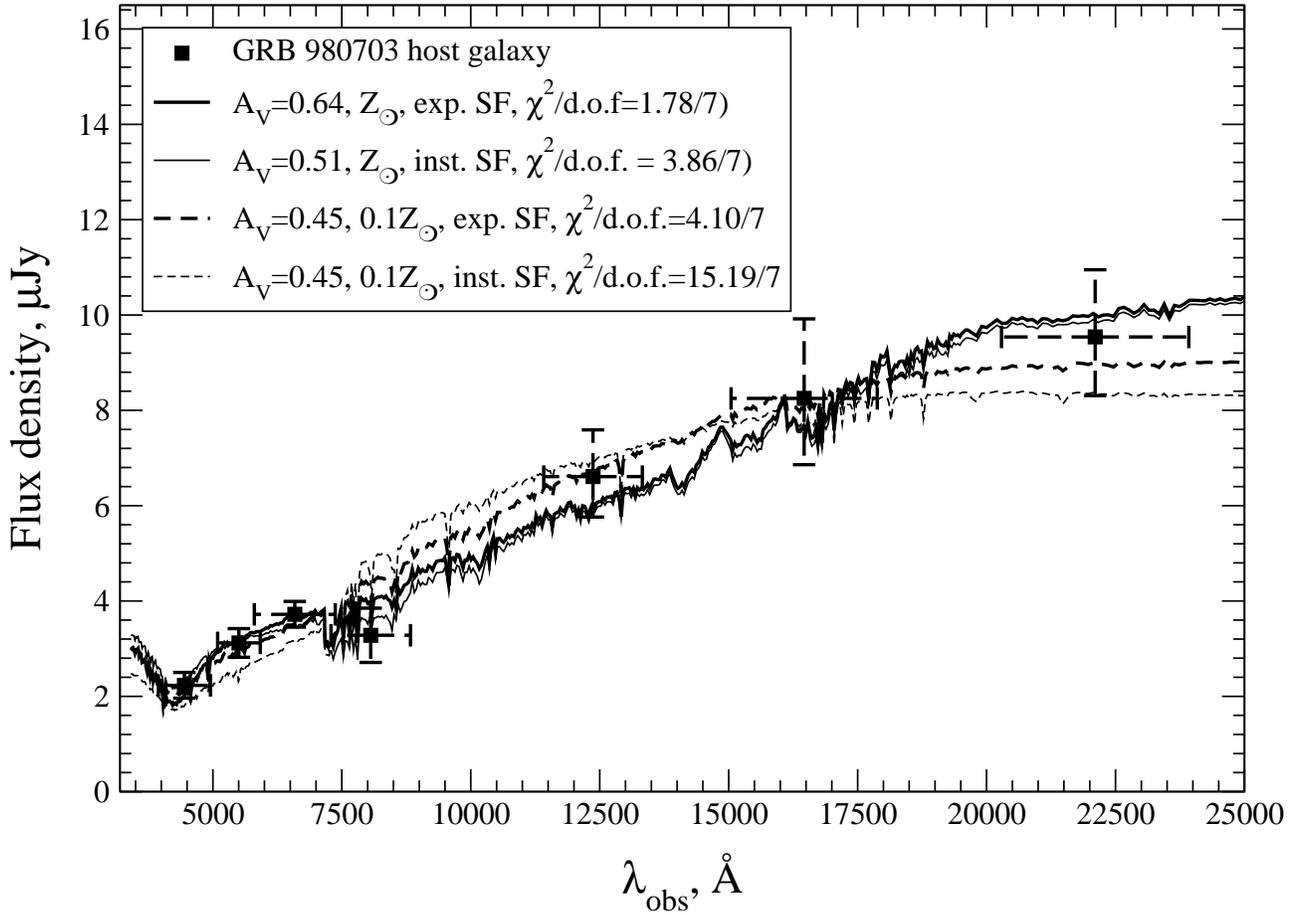}
\caption{The set of the model template SEDs for the host galaxy of GRB~980703
assuming the Cardelli et al.
extinction law. The $JHK$ data are taken from Bloom et al. (\cite{Bloom98a}).
The observed wavelengths are given.}
\label{GRB980703_MODELS}
\end{figure*}

\begin{figure}
\resizebox{\hsize}{!}{\includegraphics[clip]{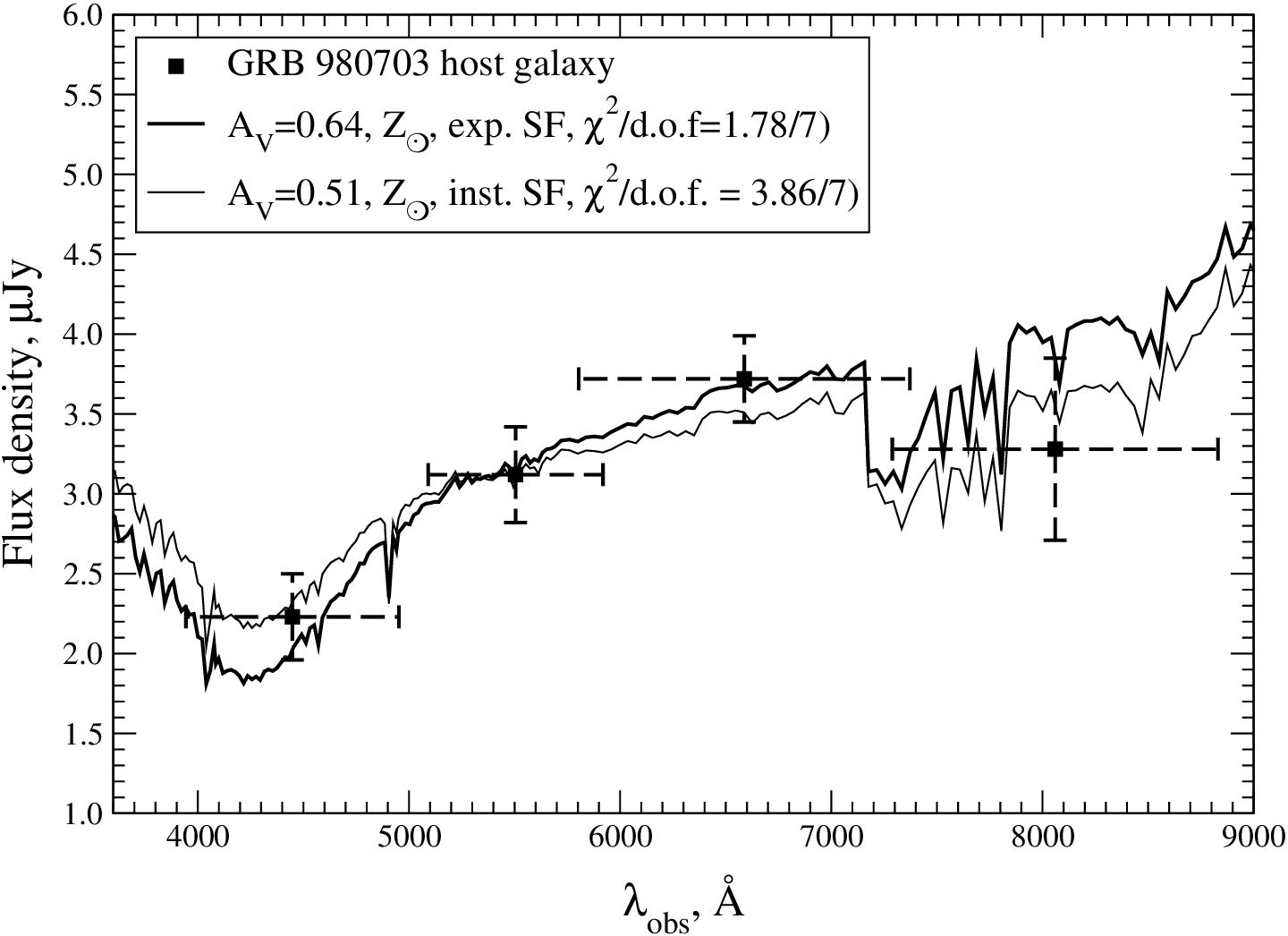}}
\caption{The SED model best fitting of $BVR_\mathrm{c}I_\mathrm{c}JHK$
data, as well as in Fig.~\ref{GRB980703_MODELS}, but shows only
$BVR_\mathrm{c}I_\mathrm{c}$
part of the spectra.}
\label{GRB980703_MODELS_best}
\end{figure}

\begin{table*}
\centering
\caption{The parameters of the theoretical template SEDs corresponding to
minimum of $\chi^2$ for the host galaxy of GRB~980703$^*$}
\label{model_GRB980703}
\begin{tabular}{lllllrlll}
\hline
\hline
Scenario      & \multicolumn{2}{l}{old component} & \multicolumn{2}{l}{burst component}    & metallicity& [\ion{O}{ii}] model    & $\chi^2_\mathrm{min}$/d.o.f. & $A_\mathrm{V}$ \\
	      &  Age, Gyr       &  Mass, \Msun~~~  &  Age, Myr        &  Mass, \Msun       &            & flux$^{**}$ &          &       \\
\hline
Instantaneous &    3            &$3.16\cdot10^{10}$&  3               & $1.78\cdot10^8$    & $Z_\odot$  & $27.8\cdot10^{-17}$ & 3.86/7               & 0.51  \\
burst         &    1.8          &$1.78\cdot10^{10}$&  4               & $1.78\cdot10^8$    &$0.1Z_\odot$& $27.4\cdot10^{-17}$ & 15.19/7              & 0.45  \\
	      &                 &                  &                  &                    &            &                  &                      &       \\
Exponential   &6 ($\tau=350$Myr)&$3.16\cdot10^{10}$&16 ($\tau=150$Myr)& $5.62\cdot10^9$    & $Z_\odot$  & $30.7\cdot10^{-17}$ &1.78/7                & 0.64  \\
decreasing    &13 ($\tau=50$Myr)&$3.16\cdot10^{10}$&40 ($\tau=500$Myr)& $1.00\cdot10^{10}$ &$0.1Z_\odot$& $31.5\cdot10^{-17}$ &4.10/7                & 0.45  \\

\hline
\hline
\end{tabular}
\parbox[t]{\textwidth}{$^*$The parameters were obtained by fitting to
$BVR_\mathrm{c}I_\mathrm{c}JHK$ data with observed flux of [\ion{O}{ii}]
emission line,
which is\\$(30.4\pm3)\cdot10^{-17}$\,erg\,s$^{-1}$\,cm$^{-2}$ (see text).\\
$^{**}$ The flux of the [\ion{O}{ii}] emission line derived in modeling
(in units of erg\,s$^{-1}$\,cm$^{-2}$).
}
\end{table*}

\begin{table*}
\centering
\caption{The selected parameters of the two host galaxies}
\label{final_model}
\begin{tabular}{llrccrll}
\hline
\hline
Host       &  Scenario         & metallicity  & Total mass         &  Age     & $A_\mathrm{V}$  & observed SFR$^{*}$ & corrected SFR \\
\hline
GRB~970508 &  instant. burst   & $0.1Z_\odot$ & $3.48\cdot10^8$    & 160\,Myr & 1.6    & $\ge 1.4$\,\Myr    & 14\,\Myr \\
GRB~980703 &  exp. decreasing  & $Z_\odot$    & $3.72\cdot10^{10}$ & 6\,Gyr   & 0.64   & $\ge 10$\,\Myr     & 20\,\Myr \\
\hline
\hline
\end{tabular}
\parbox[t]{0.84\textwidth}{$^{*}$ The SFR was recomputed following cosmology with
$H_0$=60 km$\cdot$sec$^{-1}\cdot$Mpc$^{-1}$, $\Omega_\mathrm{M}$=0.3 and
$\Omega_\Lambda$=0.7.}
\end{table*}

\section{Estimates of the SFR for the host galaxy of GRB~991208}\label{Est_SFR}
On 13 and 14 December 1999 the spectrum of the GRB~991208 optical counterpart
was obtained with the 6-m telescope of SAO RAS (see Fig.~\ref{spec}). Four
emission lines were
detected. These were $[\ion{O}{ii}]\ 3727\mbox{\AA}$, H$\beta\ 4861\mbox{\AA}$ and
$[\ion{O}{iii}]\ 4959\mbox{\AA}, \ 5007\mbox{\AA}$ at a redshift of
$z=0.7063\pm 0.0017$
(Dodonov et al. \cite{Dodonov}). The line detection allowed us to estimate the
SFR (star-forming rate) using fluxes of the emission lines [\ion{O}{ii}] and
H$\beta$.
Table~\ref{SFR} presents the fluxes of the emission lines and
corresponding SFR. Estimators of Kennicutt (\cite{Kennicutt}) for
 [\ion{O}{ii}] and Pettini et al. (\cite{Pettini}) for  H$\beta$ were used for
estimating the SFR.

\begin{figure}
\resizebox{\hsize}{!}{\includegraphics[clip]{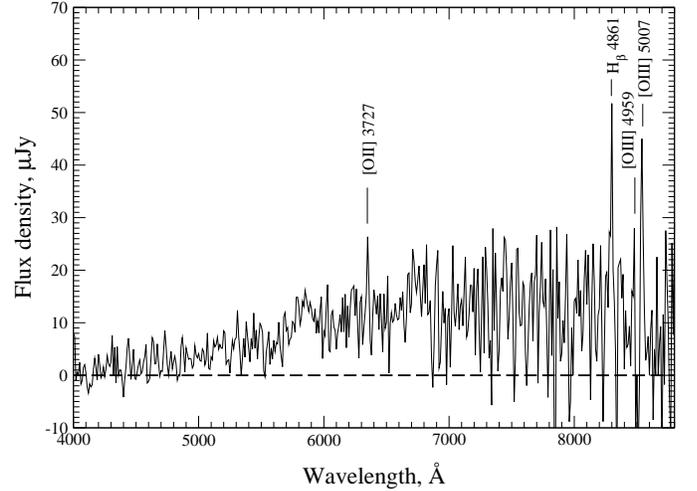}}
\caption{The spectrum of the GRB~991208 optical transient
(Dodonov et al. \cite{Dodonov}) obtained with MPFS at the SAO RAS 6-m
telescope.}
\label{spec}
\end{figure}

\begin{table*}[t]
\centering
\caption{The SFRs of the host galaxy of GRB~991208}
\label{SFR}
\begin{tabular}{lccll}
\hline
\hline
Line      &           Flux                        & observed SFR  & corrected SFR, \Myr  & corrected SFR, \Myr \\
       &  $10^{-16}\cdot \frac{erg}{cm^2\,s}$  & \Myr             & (SB, $A_\mathrm{V}\ge2.07$)   &(MW, $A_\mathrm{V}\ge1.58$) \\
\hline
$[\ion{O}{ii}]$\ 3727\AA   &      1.79$\pm$0.22   &  7.6$\pm$2.4  & $\ge$118             & $\ge$71\\
H$\beta$\ 4861\AA          &      3.84$\pm$0.33   &  28.7$\pm$1.0 & $\ge$249             & $\ge$156\\
\cline{2-5}
       &          {\bf UV}                     &{\bf continuum}& {\bf on 31 March 2000}  &\\
2800\AA\ Salpeter IMF   & 0.38$\pm$0.09 $\mu$Jy   & 1.46$\pm$0.34 & $\ge$44              & $\ge$24\\
2800\AA\ Scalo IMF  & 0.38$\pm$0.09 $\mu$Jy       & 2.26$\pm$0.54 & $\ge$68              & $\ge$38\\
\hline
\hline
\end{tabular}
\end{table*}
The $BVR_\mathrm{c}I_\mathrm{c}$ photometry obtained on 31 March 2000
for the pure host galaxy allows us to estimate the SFR in an independent
way. By interpolating between the $B$ and $V$ bands we derived the value of
the flux at 2800\AA\ in the rest frame. The flux density is
$0.38\pm 0.09\ \mu$Jy.
The use of the Madau et al. (\cite{Madau}) estimator yielded the values given
in Table~\ref{SFR}.

As can be seen from Table~\ref{SFR}, there is a difference between the
estimates due to the fact that internal extinction should be
taken into account. The comparison with local starburst galaxies has shown
that the $BVR_\mathrm{c}I_\mathrm{c}$ broad-band flux spectrum is best
fitted by the S5 average spectral energy distribution (see Sect.~\ref{BVRI-SB}).
For the S5 SED, the colour excess $E(B-V)$ is 0.51--0.60 mag.
It corresponds
to the rest-frame $V$-band extiction of $A_\mathrm{V}$=2.25$\pm$0.18 mag
and $A_\mathrm{V}$=1.72$\pm$0.14 mag
for the Calzetti et al. and Cardelli et al. reddening laws, respectively.

There is an independent way to estimate the SFR.
The observed upper limit from the OT spectrum on the H$\gamma$/H$\beta$
ratio is $\le$0.39,
which corresponds to $A_\mathrm{V}\ge$1.3 mag (Whitford \cite{Whitford}).
This lower limit is in agreement with the limit derived from the comparison
with the S5 template SED.
Here we assumed the theoretical value of H$\gamma$/H$\beta$ to be equal to
0.469$\pm$0.009 as expected for the case of B recombination.

Table~\ref{SFR} (the last two columns) gives the values of the SFR corrected
for the internal extinction,
if we assume a lower limit on $A_\mathrm{V}$=2.07 mag and
$A_\mathrm{V}$=1.58 mag for the Calzetti et al.
and Cardelli et al. reddening laws, respectively.
The SFRs have been computed
assuming Calzetti et al. (\cite{Calz2000}) and Cardelli et al. (\cite{Card})
SB and MW reddening laws, correspondently.
These values are higher by a factor of $\sim 2$ than the SFRs given by
Vreeswijk et al. \cite{Vreeswijk2000b}
for the GRB~990712 host galaxy, but in agreement within their errors.
As can be seen from Table~\ref{SFR}, the SFRs differ again. But in principle,
these estimators may give different values of the SFR.
First, the estimates from the [\ion{O}{ii}] $\lambda$3727 luminosity are sufficiently
sensitive to the density of enviroment where the emission lines form.
Second, the intensity of nebular emission lines is sensitive to the most massive
stars and these estimates give ``current" rate of star formation for
a mean time interval of a few million years.
On the other hand, the
characteristic time of averaging for estimates from the UV continuum is
about 10$^8$ years. So, comparing the different method estimates we may
trace the star formation history. Thus, from our extinction corrected
SFRs we can conclude that, most probably, there is a recent vigorous
burst of star formation in the GRB 991208 host galaxy.

\section{Discussion}\label{Disc}
We fulfilled the modeling of SEDs for the galaxies from Table~\ref{summary}
with the highest and the lowest luminosity: the host galaxy of GRB~970508 and
GRB~980703. Our main aim here was to estimate the extinction in the GRB
hosts. As was described in Sect.~\ref{Modelling}, we achieved this
goal using two approximations. First, the range of $A_\mathrm{V}$ was derived
using comparison of the $BVR_\mathrm{c}I_\mathrm{c}$ broad-band flux
spectra with local starburst templates.
The second approximation is the theoretical modeling (or computing the
theoretical templates), which allows us to choose
the reddening law and the certain value of $A_\mathrm{V}$. The selected model
parameters for these two GRB host galaxies are presented in
Table~\ref{final_model}. Let us discuss the results of the modeling.

{\it The host galaxy of GRB~970508:}
As was noted in \ref{Modeling_GRB970508}, in the case of
the host galaxy of GRB~970508 there is some uncertainty in the choice of
the reddening law. As the first approximation we derived the range of
$A_\mathrm{V}$ for both reddening laws, $A_\mathrm{V}=2.0-2.43$ mag and
$A_\mathrm{V}=1.5-1.86$ mag for Calzetti et al.
and Cardelli et al. extinction law, respectively. The modeling was performed
in these ranges and the minimum of $\chi^2$ actually occurs within both
of them.
We found that for the Cardelli et al. extinction law the
value of $A_\mathrm{V}$ does not depend on the star formation scenario and
metallicity, while for Calzetti's law this dependency is clearly seen.
As can be seen from Tables~\ref{model_GRB970508_Calz}
and \ref{model_GRB970508_Card},  in some cases the minimal value of
$\chi^2$ takes place for models with burst component masses of the same order
of old component mass or even higher. This is unlikely to be real.
It is clear that a rather large amount of gas in the galaxy is needed to
initiate formation of such a massive population. As it follows from the tables
this mass of gas is to exceed the total mass of galaxy or to be at least of
the same order, which seems to be unbelievable. On the other hand, this
situation is likely to happen in the case of primeval galaxies, i.e. the
galaxies where the first burst of star formation occurs. Nevertheless,
the observations do not confirm the formation of primeval galaxies at z=0.835,
in any case. Thus, the model of the instantaneous burst scenario and a
metallicity
of $Z_\odot$ for the Calzetti et al. extinction law and the same scenario with
both metallicities in the case of the Cardelli et al. extinction law seems to be
more reasonable.
Should one come back to the ages of the old component, it can be
noticed that the theoretical template SEDs correspond to young galaxies,
which in turn leads to the conclusion
about a lower metallicity in comparison with that of the Sun.
It follows from these considerations that SED from
Table~\ref{model_GRB970508_Card} with the following parameters:
instantaneous
burst of star formation, metallicity of $0.1Z_\odot$ and
$A_\mathrm{V}=1.6$ mag
for the Cardelli et al. (MW) extinction law, is the most appropriate model.
The selected parameters of the host galaxy are given in
Table~\ref{final_model}.

It may be noted that the ratio of the
emission lines of the host galaxy is [\ion{Ne}{iii}/[\ion{O}{ii}]=0.44$\pm$0.05
(Bloom et al. \cite{Bloom98b}), which is similar to the Seyfert~2 (Sy2) ratio
of about 0.41 from Storchi-Bergmann et al. (\cite{Storchi-Bergmann}).
For these reasons, it seems that the population synthesis models cannot
describe the spectrum of the host
due to nonthermal emission from an AGN. But Fruchter et al.
(\cite{Fruchter1999b}) noted
that for two reasons, i.e. occasional location of OTs against the host galaxy
without any trend to its core and
unusually blue optical-to-infrared colours of other GRB hosts, a more
certain explanation of the
high [\ion{Ne}{iii}]/[\ion{O}{ii}] ratio would be the existence of active star 
formation in the host rather than the AGN interpretation
(see Fruchter et al. \cite{Fruchter1999b} and references therein).
Moreover, as was concluded by Vreeswijk et al.
(\cite{Vreeswijk2000b}), in the case of another GRB host galaxy, the host galaxy
of GRB~990712,
the ratio of emission lines indicates that the host
is most likely an \ion{H}{ii} galaxy, contrary to the Seyfert~2 explanation
of Hjorth et al. (\cite{Hjorth2000}).
Finally, as was discussed by Sokolov et al. (\cite{Sokolov1999}), the
GRB~970508 host would
be a blue compact galaxy with a higher surface brightness. Their Fig.~6
showing the absolute
magnitudes $M_\mathrm{B}$ vs. the linear diameter $\log D_{25}$  revealed that
the observed parameters ($M_\mathrm{B}$ and $\log D_{25}$) of the host galaxy of
GRB~970508 lie on the branch of galaxies with surface brightness higher than
normal (as it should be for Mrk galaxies), which again implies a starburst
activity.
Thus, we may conclude that the host galaxy of
GRB~970508 is most probably an actively star-forming young galaxy, which is
in agreement with the results of the modeling.
Our value of the intrisic extinction allows us to derive the corrected SFR.
Using the Cardelli et al. (\cite{Card}) reddening curve and
$A_\mathrm{V}=1.6$ mag we computed that the SFR
should be corrected by a factor of about 10. The lower limit of the SFR
from the luminosity of the [\ion{O}{ii}] emission line derived by
Bloom et al. (\cite{Bloom98b}) is about 1\,\Myr, the extinction corrected
SFR will then be about 10\,\Myr for the cosmological parameters assumed by
Bloom et al. (\cite{Bloom98b}).

{\it The host galaxy of GRB~980703:}
As was shown in Sect.~\ref{Modeling_GRB980703}, in the case of the host
galaxy of GRB~980703 the observational data make it possible to choose the
only extinction law similar to that in the Milky Way (Cardelli et al. \cite{Card}).
Thus, in the second approximation we investigated only one range of
$A_\mathrm{V}=0.3-0.65$ mag, which corresponds to the Cardelli et al.
extinction law. As can be seen from
Fig.~\ref{GRB980703_MODELS} the theoretical templates with a metallicity
of $0.1Z_\odot$ systematically deviate from the observed
$BVR_\mathrm{c}I_\mathrm{c}JHK$
broad-band fluxes, particularly in the $BVR_\mathrm{c}I_\mathrm{c}$
and $K$-band domain. The fact
that the $BVR_\mathrm{c}I_\mathrm{c}JHK$ data are best described by
the models with
solar metallicity is naturally explained by the age of the old component
of the stellar population. Indeed, a period of time of order of 6\,Gyr is enough
to enrich the medium with heavy elements.
The final parameters of the host galaxy are given in
Table~\ref{final_model}.
As in the case of the host galaxy
of GRB~970508, if the internal extinction is known, we can estimate
the SFR corrected for the latter.
As was concluded by Djorgovski et al. (\cite{Djorgovski}), the lower limit
to the SFR for the GRB~980703 host is
about 7\,\Myr, which was derived from the $H_\beta$ luminosity. Then, using the
Cardelli et al. (\cite{Card}) reddening law and $A_\mathrm{V}=0.64$ mag,
we derive the SFR to be $\sim$14\,\Myr for the cosmology assumed in
Djorgovski et al. (\cite{Djorgovski}).}

To summarize the modeling, we emphasize that
due to the cosmological redshift, our $BVR_\mathrm{c}I_\mathrm{c}$
photometry for the GRB
host galaxies covers only the ultraviolet--blue
part of the spectral range, which is too short. For a more certain modeling,
as was demonstrated in the case of the host of GRB~980703,
we need UV and IR observations of hosts, but this is
a nontrivial task for objects with such low observed fluxes. But, as was
shown, our photometric points are in excellent agreement with the spectra of the
host galaxies and it suggests that our modeling is good enough to describe the
hosts. In addition, we should note that the $BVR_\mathrm{c}I_\mathrm{c}$
spectral range (in the rest frame) is sufficiently sensitive to some features
of extinction laws. As we demonstrated in Sect.~\ref{Modelling}, in principle,
it is possible even to distiguish between different reddening laws.

For the eight GRB host galaxies from Table~\ref{hosts}
there were enough observational data for the correct estimates
of the absolute rest-frame magnitudes in $B$-band
(see Table~\ref{summary}).
Thus it would be interesting to compare the derived luminosities
of the host galaxies (Table~\ref{summary}) with
$L_*$ (where $L_*$ corresponds to the luminosity of a typical galaxy at the
``knee" of the local galaxy luminosity function). As
was discussed in Pian et al. (\cite{Pian}) an $L_*$ galaxy has
$M_{\mathrm{B}_*}\approx-21$. Using a cosmological model with
$H_0$=60\,km$\cdot$sec$^{-1}\cdot$Mpc$^{-1}$, $\Omega_\mathrm{M}$=0.3 and
$\Omega_\Lambda$=0.7 as in Sect.~\ref{Est-Mabs},
we estimate the absolute magnitudes presented in Table~\ref{summary},
which have typical errors of $M_{\mathrm{B}_\mathrm{rest}}$
from 0.3 to 0.4 mag
derived from the calculations of the  $K$-correction in Sect.~\ref{Est-Mabs}.
\begin{table*}
\caption{The observed and absolute magnitudes of the GRB host galaxies }
\label{summary}
\centering
\begin{tabular}{llll}
\hline
\hline
Host       &  observed magn. &  absolute magn. &  reference \\
	&   $R$           &  $M_{\mathrm{B}_\mathrm{rest}}$ &            \\
\hline
GRB~970228 &  24.6$\pm$0.2   &  -18.6  & $R$: Galama et al. (\cite{Galama}) \\
	&                 &            & $M_{\mathrm{B}_\mathrm{rest}}$: Bloom et al. (\cite{Bloom2000b})\\
GRB~970508 &  24.99$\pm$0.17 &  -18.6  & This paper \\
GRB~971214 &  25.69$\pm$0.3  &  -21.1  & $R$: This paper \\
	&                 &            & $M_{\mathrm{B}_\mathrm{rest}}$: Kulkarni et al. (\cite{Kulkarni})\\
GRB~980613 &  23.58$\pm$0.1  &  -20.8  & This paper \\
GRB~980703 &  22.30$\pm$0.08 &  -21.3  & This paper \\
GRB~990123 &  24.47$\pm$0.14 &  -20.9  & This paper \\
GRB~990712 &  21.80$\pm$0.06 &  -19.9  & Hjorth et al. \cite{Hjorth2000}\\
GRB~991208 &  24.36$\pm$0.15 &  -18.8  & This paper \\
\hline
\hline
\end{tabular}
\end{table*}
The values of $M_{\mathrm{B}_\mathrm{rest}}$ given by other authors were
recomputed according
to the cosmology assumed by us. However, these magnitudes were computed
in a different way.
Bloom et al. (\cite{Bloom2000b}) computed the absolute magnitude from the
observed continuum in the rest frame.
Kulkarni et al. (\cite{Kulkarni}) assumed an intrisic
host galaxy spectrum with $F_\nu$ $\propto\,\nu^{-0.7}$ and extrapolating
from the observed $R$-band magnitude.
They computed  $M_{\mathrm{B}_\mathrm{rest}}$ from the continuum of the
host galaxy spectrum.
In the case of the host galaxy of GRB~990712,
$M_{\mathrm{B}_\mathrm{rest}}$ was derived by Hjorth et al. (\cite{Hjorth2000})
assuming that
for $z=0.434$ the observed $R$-band is approximately equivalent to the
rest-frame $B$-band.
Note that our estimates were made by calculating the $K$-correction.
Comparing $M_{\mathrm{B}_*}$ and $M_{\mathrm{B}_\mathrm{rest}}$ from
Table~\ref{summary}, we can
conclude that the luminosities of the host galaxies of GRB~970228, GRB~970508
and GRB~991208 are $\sim$ 0.1 $L_*$, but the hosts of GRB~980613,
GRB~980703 and GRB~990123 have values close to $L_*$.
At least, it seems that the luminosities of the hosts are certainly not
above the ``knee" of the local luminosity function.

We have shown that there is a significant
internal extinction in these GRB host galaxies. The comparison with local
star-forming SEDs and our modeling demonstrate the influence of the
reddening. Taking into account the results of the modeling, the estimates of
rest-frame extinction in the $B$-band for the host galaxies of GRB~970508 and
GRB~980703 yield  $A_\mathrm{B}=2.14$ and $A_\mathrm{B}=0.86$ magnitudes,
respectively. Then, the corrected absolute magnitudes for these hosts are
$M_{\mathrm{B}_\mathrm{rest}}\approx-20.7$ and $M_{\mathrm{B}_\mathrm{rest}}\approx-22.0$,
respectively.
These estimates are about or $\sim$ 1 mag higher than $M_{\mathrm{B}_*}$.
Perhaps, the
observational fact that GRB host galaxies are underluminous is due to
obscuring of the UV and optical flux by dust and/or dense molecular clouds.

It would be interesting
to compare our $M_{\mathrm{B}_\mathrm{rest}}$ magnitudes from
Table~\ref{summary} to a recent paper by Fried et al. (\cite{Fried}).
These authors present results from the CADIS survey which
includes a study of the luminosity function of various types
of galaxies as a function of redshift (in particular their Table~2).
They present the data for an $H_0=100$\,km\,s$^{-1}$\,Mpc$^{-1}$,
$\Omega_\Lambda = 0.7$ and $\Omega_\mathrm{M} = 0.3$ cosmological model.
To transform the $M^{*}$ magnitudes from Table~2 of  Fried et al. (\cite{Fried})
to our cosmology (with  $H_0=60$\,km\,s$^{-1}$\,Mpc$^{-1}$)
we used the relation $M_{60} = M_{100}-1.1$. Then
the $M^{*}$ magnitudes are increased to be $M^{*}\approx -20.1\pm0.2$ mag in
the redshift range of $0.3-0.5$, $M^{*}\approx -21.5\pm0.3$ mag in the redshift
range of $0.5-0.75$ and $M^{*}\approx -21.0\pm0.2$ mag in the redshift range of
$0.75-1.04$ for starburst galaxies from Table~2 by Fried et al. (\cite{Fried}).
The comparison with the absolute magnitudes from our
Table~\ref{summary} shows that the $M_{\mathrm{B}_\mathrm{rest}}$ of the host
galaxies of GRB~990712, GRB~980613 and GRB~980703 corresponds to $M^{*}$ from
Fried et al. (\cite{Fried}). The $M_{\mathrm{B}_\mathrm{rest}}$ of the hosts of
GRB~970228, GRB~970508 and GRB~991208 are much lower than the $M^{*}$, but
we pay attention to the fact that the observed $I$-band magnitudes of these
hosts are fainter by 1 mag than the completeness limit of the CADIS survey
($I_{815}=23.0$, Fried et al. \cite{Fried}). The redshifts of other two hosts
from our Table~\ref{summary} are out of the redshift range of the CADIS survey.

\section{Conclusions}\label{Concl}
To summarize, we can conclude the following.

(i) There is a significant extinction in the host galaxies
(up to $A_\mathrm{V}\sim2$ mag).
It is important to take into account effects of the reddening in the
estimation of SFR. The GRB host galaxies show a high rate of star-formation
if we correct the observed values for internal extinction, from tens to hundreds
of solar masses per year.

(ii)
Two host galaxies (GRB~970508 and GRB~991208)
have equal luminosities and are fitted by the same local starburst template
SEDs (S5), which means that these hosts have very similar internal extinction.
For both of them the SFRs turn out to be high, and for the host galaxy
of GRB~991208 the SFR is the highest (more that 150--200 \Myr) from all
known GRB hosts, judging from fluxes in emission lines of this host galaxy.

(iii) In the case of the host galaxy of GRB~980703 the observed deficit in
the $B$-band can be explained by the excess of extinction near
$2200\mbox{\AA}$, which is characteristic of the extinction law similar to
that of the Milky Way.

(iv) There is a connection between GRB and \ion{H}{ii} regions
(see Sect.~\ref{BVRI-SB}, Ahn \cite{Ahn}), and together with
the association of the host galaxy of GRB~990712 with \ion{H}{ii} galaxies
(Vreeswijk et al. \cite{Vreeswijk2000b}), again implies strong evidence
for massive star formation.

Thus, we conclude that long-duration GRBs seem to be closely related to
vigorous massive star-forming in their host galaxies. It should be noted
that the SFR in the host galaxies is unlikely to be much higher than
in galaxies at the same redshifts ($z\sim1$). As was shown by
Glazebrook et al. (\cite{Glazebrook}) at this redshift the mean star formation rate
is about 20-60\,\Myr (see also Blain \& Natarajan \cite{Blain}). For these
reasons we conclude that GRB host galaxies
seem to be similar to field galaxies at the same redshift.

\begin{acknowledgements}
We thank S.G. Djorgovski and J.S. Bloom for kind permission to use
two figures (Figure~2 and Figure~3) from their papers.
We thank V.I. Korchagin and Yu.V. Baryshev for fruitful discussion of
the results. We also thank our referee (S. Klose) for useful comments
that improved our manuscript on several important points.
The work was supported by the INTAS N96-0315, ``Astronomy"
Foundation (grant 97/1.2.6.4) and RFBR N98-02-16542, RFBR N00-02-17689.
\end{acknowledgements}

\end{document}